\documentclass{article}

\usepackage[UKenglish]{babel}
\usepackage{graphicx} 
\usepackage{microtype}
\usepackage[dvipsnames]{xcolor}
\usepackage{graphics}
\usepackage{booktabs}

\usepackage{amsmath}
\usepackage[colorlinks=true, allcolors=blue]{hyperref}
\usepackage{afterpage}
\usepackage[official]{eurosym}
\usepackage{hyperref}
\usepackage{placeins}
\usepackage{authblk}
\usepackage[authoryear]{natbib}
\usepackage[utf8]{inputenc}
\usepackage{tikz}
\usepackage[flushleft]{threeparttable}
\usepackage{lscape}
\usepackage{adjustbox}
\usepackage{copyrightbox}
\usepackage{float}
\usepackage[position=top]{subfig}
\usepackage{multirow}
\usepackage{longtable}
\usepackage{caption}
\usepackage{threeparttablex}
\usepackage{ltablex}
\usepackage{geometry}

\newcommand{\sym}[1]{\ifmmode^{#1}\else\(^{#1}\)\fi}

\title{Modelling Distributional Impacts of Carbon Taxation: a Systematic Review and Meta-Analysis}
\newcommand\correspondingauthor{\thanks{Corresponding author.Email address: jules.linden@liser.lu (J. Linden)}}

\author[1]{Jules Linden\correspondingauthor}
\affil[1]{Luxembourg Institute of Socioeconomic Research
}

\author[2]{Cathal O'Donoghue}

\author[1]{Denisa M. Sologon}

\affil[2]{University of Galway
}

\begin{document}
\maketitle

\begin{abstract}
Carbon taxes are increasingly popular among policymakers but remain politically contentious. A key challenge relates to their distributional impacts; the extent to which tax burdens differ across population groups. As a response, a growing number of studies analyse their distributional impact ex-ante, commonly relying on microsimulation models. However, distributional impact estimates differ across models due to differences in simulated tax designs, assumptions, modelled components, data sources, and outcome metrics. This study comprehensively reviews methodological choices made in constructing microsimulation models designed to simulate the impacts of carbon taxation and discusses how these choices affect the interpretation of results. It conducts a meta-analysis to assess the influence of modelling choices on distributional impact estimates by estimating a probit model on a sample of 217 estimates across 71 countries. The literature review highlights substantial diversity in modelling choices, with no standard practice emerging. The meta-analysis shows that studies modelling carbon taxes on imported emissions are significantly less likely to find regressive results, while indirect emission coverage has ambiguous effects on regressivity, suggesting that a carbon border adjustment mechanism may reduce carbon tax regressivity. Further, we find that estimates using older datasets, using explicit tax progressivity or income inequality measures, and accounting for household behaviour are associated with a lower likelihood of finding regressive estimates, while the inclusion of general equilibrium effects increases this likelihood. 

\end{abstract}

\section{Introduction}

Carbon taxation is regarded as one of the most efficient policy instruments for reducing greenhouse gas emissions. Yet, carbon prices remain too low to effectively mitigate climate change \citep{agnolucci2024measuring, WB2024StateofCarbonPricin}. Low carbon prices are partly due to low public support, driven by concerns about their cost of living and distributional impacts \citep{douenne2020french, dechezlepretre2022fighting}. A growing literature analyses the distributional impacts of carbon taxation ex-ante, frequently utilizing environmental microsimulation models. These models use micro-level data to simulate policy effects at the individual or household level, accounting for population heterogeneity \citep{o2014handbook}. Despite an expanding literature, findings on carbon tax regressivity remain inconsistent and qualitative conclusions on carbon taxation's distributional impact in the same country can sometimes differ across studies\footnote{France: regressive \citep{berry2019distributional, douenne2020vertical, giraudet2021policies, bourgeois2021lump}, progressive \citep{feindt2021understanding}; Italy: regressive \citep{vandyck2021climate}, proportional \citep{tiezzi2005welfare}; Mexico: regressive \citep{renner2018household, rosas2017distributional}, progressive \citep{renner2018poverty}; Poland: regressive \citep{Antosiewicz22}, progressive \citep{feindt2021understanding}; Spain: regressive \citep{vandyck2021climate, symons2002distributional}, proportional \citep{labandeira1999combining}; UK: regressive \citep{symons1994carbon, barker1998equity}, proportional \citep{symons2002distributional}; USA: regressive \citep{hassett2009incidence, mathur2014distributional, grainger2010pays}, proportional \citep{rausch2011distributional}}. 

This paper examines how modelling choices shape distributional impact estimates and helps explain why studies analysing similar carbon tax policies sometimes reach divergent conclusions. We first identify the main drivers of carbon taxation's distributional impact, document how these drivers are implemented in environmental microsimulation models, and quantify how alternative ways of modelling these drivers affect reported outcomes. While previous work has shown that high-level modelling choices, such as the inclusion of behavioural responses, influence distributional impact estimates \citep{ohlendorf2021distributional}, the role of more granular implementation choices, such as the data base used, remains less well understood. Because constructing environmental microsimulation models involve numerous interrelated modelling and implementation decisions, estimates can vary substantially across studies, complicating interpretation and cross-study comparison. In taking stock of modelling approaches and assessing their impact on model estimates, this paper supports researchers in developing models and helps readers interpret and compare study results. 



This review focuses on micro-simulation models designed to assess the distributional impacts of carbon taxation. Existing review articles synthesize insights from the literature on carbon taxation, including on their effectiveness, impacts on competitiveness and innovation, public acceptance, and their distributional impacts \citep{wang2016distributional, Köppl22, tietenberg2013reflections, markkanen2019social, cuevas2024health, speck1999energy, timilsina2022carbon, marron2014tax, pizer2019distributional, shang2023poverty}. \cite{wang2016distributional} review ex-ante and ex-post studies on distributional impacts, distinguishing between the impact across household groups and sectors. \cite{shang2023poverty} focus on the poverty and distributional impacts of carbon pricing, differentiating between the consumption, income, health, and revenue recycling channels.  \cite{pizer2019distributional} focus on distributional impacts of energy taxes and horizontal equity impacts. \cite{markkanen2019social} include economic, health, gender, and ethnic equality impacts of various climate change mitigation measures, while \cite{cuevas2024health} focuses on the health impacts. 

Despite a growing literature, few review articles focus on methodological approaches. \cite{timilsina2022carbon} discuss General equilibrium (GE) and Input-Output (IO) models but omit ex-ante microsimulation models. \cite{ohlendorf2021distributional}'s meta-analysis provides the most advanced treatment of methodological approaches. Pooling estimates across methodologies, their study quantifies the impact of high-level modelling choices (GE effects, household behaviour, tax design, indirect effects, and the choice of welfare concept) on the likelihood of finding regressive distributional impacts, but it does not consider how these choices are implemented. They find that studies on the transport sector, those incorporating demand-side responses and indirect effects, and those using lifetime income proxies are more likely to report progressive results. Compared to \cite{ohlendorf2021distributional}, our study narrows the focus to studies utilizing microsimulation modelling and expands on the determinants considered, distinguishing for the first time between conceptual modelling choices (e.g. whether to model indirect emissions) and their implementation (e.g. whether to use a single region or multi-regional IO model to model indirect emissions). 

This paper makes two contributions to the literature. First, it provides an in-depth review of the components of microsimulation models designed to assess carbon taxation and describes how model components have been implemented. Second, it estimates the impact of modelling choices and their implementation on the likelihood of finding regressive distributional impacts, going beyond the factors considered previously. 

Using a sample of 217 estimates across 71 countries, this paper’s meta-analysis demonstrates that implementation choices significantly impact distributional impact estimates and help explain variation across studies. Expanding on \cite{ohlendorf2021distributional}, we find that including implementation-related predictors increases the explanatory power of a probit model by  55\%. A key insight from our analysis is that \cite{ohlendorf2021distributional}'s observed link between indirect effects and progressive estimates is primarily driven by imported emissions rather than indirect effects per se. This result suggests that expanding tax coverage to emissions embedded in imported goods, through a Carbon Border Adjustment Mechanism for example, reduces carbon tax regressivity. Further research is, however, needed to verify this. Further, our meta-analysis shows that relying on older datasets, using explicit tax progressivity or income inequality measures, and accounting for household behaviour increases the likelihood of progressive estimates. Further, we find that the inclusion of GE effects increases the likelihood of finding regressive impacts once we account for the choice of Input-Output database and the coverage of imported emissions. 

Section \ref{method} introduces microsimulation modelling, important modelling steps, and the literature review strategy. Section \ref{choices_implementation} briefly discusses conceptual modelling choices and their implementation. For each conceptual modelling choice, we first present central insights from the literature and then discuss how the conceptual modelling choices have been implemented in the literature. Section \ref{metaanalysis} presents the results of the meta-analysis.

\section{Methodology}\label{method}

\subsection{Microsimulation modelling of carbon taxation}
Microsimulation models are simulation-based tools with a micro-unit of analysis used to analyse the impact of policy, economic, or social changes ex-ante \citep{o2021practical}. By simulating policy designs on micro-level data, these models inform on the complex distributional outcomes resulting from the interaction of policy design (tax coverage or revenue recycling) with existing policies and a heterogeneous population. 


A primary advantage of microsimulation models is their ability to account for heterogeneous populations. Two sources of heterogeneity are particularly relevant for modelling carbon taxes: heterogeneity in incomes and preferences, resulting in heterogeneous consumption patterns and behavioural responses to price changes.  

Environmental microsimulation models, frequently used to assess distributional impacts of carbon taxation, are similar to an indirect tax model \citep{hynes2014environmental, decoster2011microsimulation}. They use household-level expenditure data, a policy calculator, and sometimes include behavioural responses at the intensive margin \citep{creedy2006carbon, reanos2022measuring}. Environmental taxes are commonly levied in relation to quantities, while most indirect (and direct) taxes are levied in relation to value. The main additional component of environmental microsimulation models, therefore, relates to the modelling of quantities and associated pollution \citep{hynes2014environmental}. 


These models can be linked to macroeconomic models to compute indirect emissions \citep{Kitzes2013} or model changes in firm behaviour and the resulting impacts on employment and incomes \citep{vandyck2021climate, Antosiewicz22}. Sometimes, environmental microsimulation models are combined with other datasets and tax-benefit microsimulation models to model revenue recycling schemes \citep{maier2024carbon, akouguz2020new}. Some microsimulation models were complemented with specialized models, scenarios, and data sets to model household behaviour at the extensive margin \citep{bourgeois2021lump, jacobs2022distributional}.

\subsection{Theoretical overview and study focus}
\begin{figure}[ht]
	\centering
		\caption{Modelling steps and choices.}

		\copyrightbox[]{\includegraphics[width=1\textwidth]{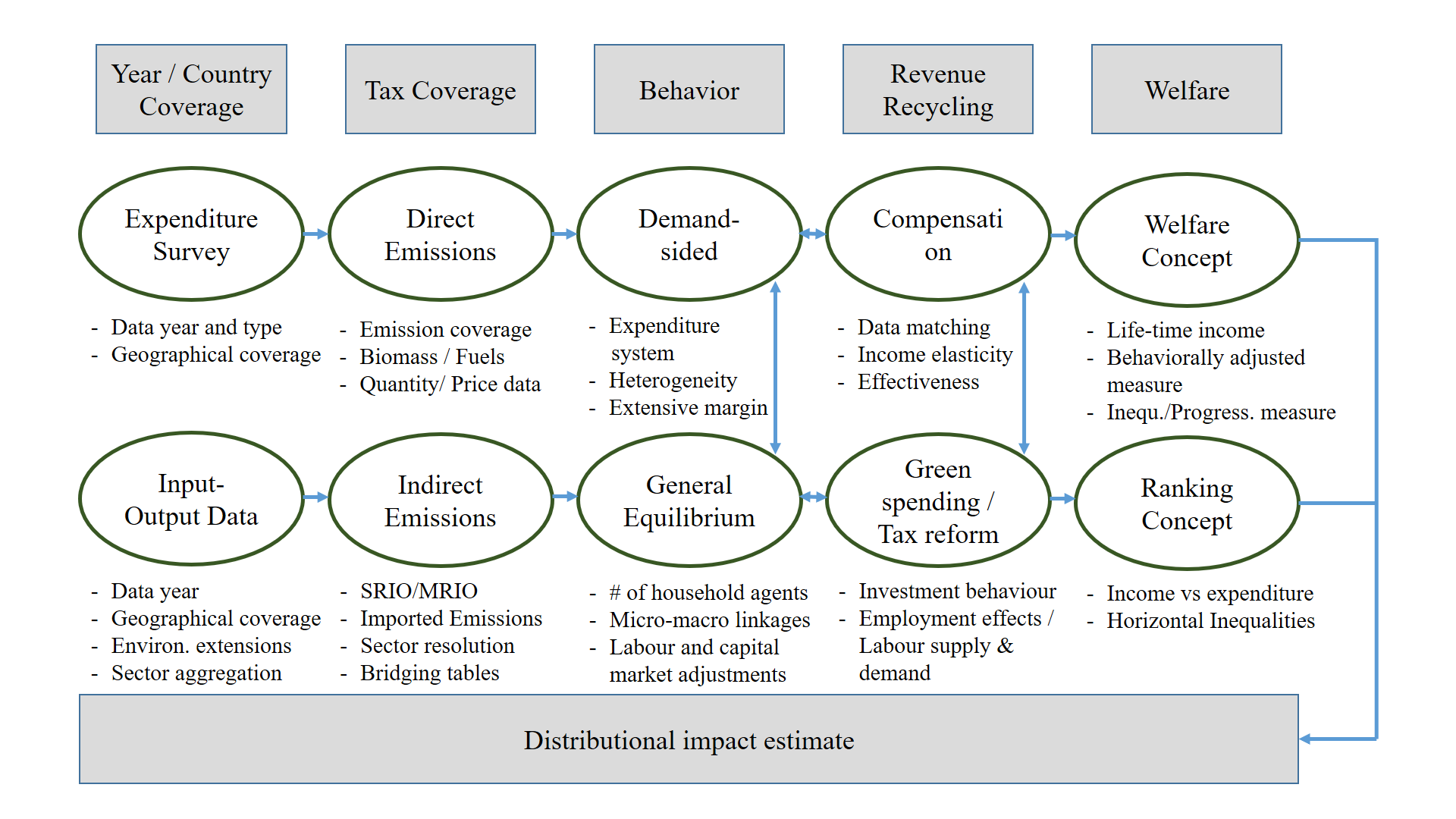}}%
		{}
		
	\label{FIG:TheoryFramework}
\end{figure}

Figure \ref{FIG:TheoryFramework} summarizes the steps and choices involved in constructing environmental microsimulation models, distinguishing between broad modelling choices (in boxes), related conceptual choices (in circles), and their implementation (in bullet points).

A first choice concerns data and geographical coverage. Expenditure surveys serve as primary datasets and contain information on households' purchases, demographic characteristics, and income. They are, however, not always published at consistent frequencies and may be severely outdated. Input-output data is needed to compute indirect emissions associated with the production and transportation of goods and services. While various country-specific and multi-regional IO models are available, they may also be outdated or unavailable in some countries. 

The next choice concerns tax coverage, including the emission type (direct, indirect, $CO_2$ or $CH_4$) and the sectors covered. In modelling direct emissions, the granularity of the expenditure, quantity, and price data may determine the level of differentiation between energy commodities. Impacts on non-energy goods can be modeled using environmentally-extended IO models. Available IO datasets differ in geographical, sectoral, and environmental impact coverage, and may restrict the policy designs that can be studied. For example, Multi-Regional IO (MRIO) models allow estimating the impact of a Carbon Border Adjustment Mechanism, while this is not possible with Single-Region IO (SRIO) models. 

Further, in developing models, researchers decide whether to model household and firm behaviour. Relevant choices relate to the way household behavioural responses are estimated, the heterogeneity across agents, and the method to link micro- and macro-models. When researchers are interested in assessing the impact of household behaviour in a partial equilibrium setting, they will need to choose how to integrate price elasticities, either through the estimation of expenditure systems or by 'borrowing' estimates from elsewhere. When microsimulation and General equilibrium models are combined, modellers decide on how to link these models and how to model adjustments in the labour (e.g. assume full employment or incorporate labour market frictions) and capital markets (modelling changes in capital incomes uniformly, or differentiated by sector or asset class). 

Carbon prices raise revenues, which can fund compensation schemes, tax cuts, and green spending. Modelling such reforms can require matching expenditure surveys with other datasets. Estimating the environmental impacts of these reforms requires estimating income elasticities or technology adoption (changes along the extensive margin) and the likelihood of transitioning into (un)employment.

Lastly, computing the carbon tax burdens requires a definition of welfare (a welfare concept), and comparing the impacts across groups requires some definition of (income) groups (a ranking concept). 

The implications of these choices for the distributional impact estimates are the subject of this paper.

\subsection{Systematic Review}
We follow the PRISMA (Preferred Reporting Items for Systematic Reviews and Meta-Analysis) approach to select papers to review. First, we define a search strategy and search terms. Second, we identify papers fulfilling the search criteria. Third, we screen and filter for inclusion. 

Our search strategy has three components. First, we searched the multi-disciplinary database Scopus\footnote{The search was conducted in April 2024 using the query (TITLE-ABS-KEY(carbon AND tax*)) AND ALL(microsim* OR micro-sim*) AND TITLE-ABS-KEY(distribut*)).}. This search yielded 73 results. Second, we complemented the list by reviewing these papers' reference lists. This yielded an additional 53 papers, resulting in a total sample of 126 papers. 

We apply multiple exclusion criteria, excluding 51 papers, so the final sample consists of 75 papers and 217 estimates. 

We focus on bottom-up estimates of the distributional impact of carbon taxation, excluding papers that use GE models with only a few representative agents\footnote{We exclude \cite{EKINS20112472, goulder2019impacts, telaye2019exploring, de2023impacts, gonzalez2012distributional, beck2015carbon, landis2021between, fragkos2021equity, williams2015initial, wu2022distributional, garaffa2021distributional}.}, Integrated Assessment Models \citep{zhao2022poverty}, or IO models without a link to micro-level datasets \citep{grottera2017impacts, da2016distributional}. We include GE studies with explicit top-down links to microsimulation models or with many representative agents \citep{landis2019cost, vandyck2021climate, ravigne2022fair,rausch2011distributional}\footnote{For example \cite{yusuf2015distributional}, who differentiate between 100 rural and 100 urban households.}. We exclude Agent-based models (ABM) by specifying 'microsim*' in the search term. While ABMs are also bottom-up models, they are designed to model the interactions between autonomous agents and analyse aggregate outcomes that emerge from micro-level dynamics. The ABM methodology, therefore, differs substantially from microsimulation. 

We focus our search on studies of carbon taxation, excluding studies on Emission Trading Schemes, energy taxation, subsidy reforms, and other energy policies.  

We limit our search to peer-reviewed papers in English, though some reports are discussed (for example \cite{flues2015distributional} and \cite{immervoll2023pays})\footnote{We exclude reports such as \cite{pearson1991european, feng2023global, flues2015distributional, kanzig2023unequal, immervoll2023pays, gough2011distribution, timilsina2023distributional, timilsina2024economic}}. We also exclude papers that do not provide quantitative results \citep{van2016economic, kirchner2019exploiting, wang2016distributional, buchs2011bears, Köppl22} or compute carbon footprints but do not simulate a carbon tax \citep{levay2021association, buchs2024emission, farrell2017factors, ivanova17}. However, reports and papers that are formally excluded from the meta-analysis are discussed where appropriate. 

The full list of included papers and a graphical representation of the selection process are provided in Supplementary materials \ref{listofstudies} and \ref{FIG:Framework}. 

\section{Literature review - Conceptual modelling choices and their implementation} \label{choices_implementation}

\subsection{Country coverage and the unit of analysis}
Distributional impacts of carbon taxation differ across countries and regions \citep{ivanova17, dorband2019poverty}. The majority of microsimulation-based assessments focus on EU countries, North America, and China (Figure \ref{FIG:mapSingle} of the supplementary materials). In recent years, comparative studies leveraging standardized datasets are increasingly common\footnote{Examples include \citep{symons2002distributional, pearson1991european, barker1998equity, missbach2023assessing, feindt2021understanding, buchs21, dorband2019poverty, vogt2019cash, steckel2021distributional, missbach2024cash, rub2024inequality, vandyck2021climate, jaccard2021energy, linden2024manyfaces}.}, expanding the geographical coverage \citep{steckel2021distributional, dorband2019poverty, missbach2024cash, missbach2024distributional}.

Depending on the research question, it may be appropriate to focus on one region or to compare multiple regions. For example, single-country studies often focus on the heterogeneity of burden across households or individuals, exploring specific channels, such as the role of emission coverage \citep{kerkhof2008taxation, renner2018household}, employment effects \citep{Antosiewicz22}, household behaviour \citep{west2017estimates}, or revenue recycling \citep{fremstad2019impact, berry2019distributional}, or simulating proposed or existing taxes \citep{IMMERVOLL2025114783, elgouacem2024pays}. The majority of the studies are single-country studies, and compare the impact of carbon taxes across income or expenditure groups within individual countries.

Other questions require that results be compared across countries. Here, we differentiate between two muéti-country literatures. 

The first literature examines how country characteristics shape distributional outcomes of carbon pricing, using the country as a central unit of analysis\citep{missbach2024distributional, dorband2019poverty, feindt2021understanding}\footnote{Relevant country characteristics include average income levels \citep{ivanova17}, population density and car ownership \citep{minx2013carbon}, average household size \citep{ivanova17, levay2021association}, access to infrastructure \citep{tukker2010impacts}, income inequality \citep{andersson2020distributional}, carbon-intensive value chains \citep{sager2019income, feindt2021understanding}, district heating availability \citep{kerkhof2009determinants, linden2024manyfaces}, the housing stock, carbon-intensity of the electricity supply \cite{PaisMagalhaes2020HouseholdsEC}, and climate \citep{wiedenhofer2017unequal, ivanova17}.}. A prominent insight from this literature is that carbon taxes are more likely to be regressive in richer countries \citep{ohlendorf2021distributional, dorband2019poverty}. This is explained by an inverse U-shaped relationship between energy expenditure shares and income. \cite{dorband2019poverty} identify a threshold of average incomes below \$US 15000 per year (PPP-adjusted) beyond which carbon tax progressivity is more likely. This insight has also been challenged. For example, \cite{wang2016distributional}'s literature review does not support the perception that distributional impacts are more likely to be progressive in developing countries.
 

The second comparative literature focuses on household-level determinants, often analysing the differences in the relationship between energy consumption and income across countries. Studying 8 Asian countries, \cite{steckel2021distributional} highlight the role of households' energy mix in determining the impacts, with large differences in the energy mix across income groups and countries. Comparing 6 EU countries, \cite{linden2024manyfaces} show that differences in households' (energy) purchases largely explain carbon tax regressivity in some Eastern European countries, but less so in Western European countries where differences in savings rates play are larger role. Studying 23 EU countries, \cite{feindt2021understanding} find higher contributions of indirect emissions to household tax burdens in wealthier Western European countries. In their study of 16 Latin American and Caribbean nations, \cite{missbach2024cash} show that differences in car ownership and cooking fuel use explain cross-country differences in distributional impacts.

\subsubsection{Implementation: Data Quality, Timeliness, and study scope}

Country selection and scope are often guided by data availability, restricted by high model development costs, and guided by the research question. 

While household expenditure surveys are available worldwide (they form the basis of consumer price index calculations), their publication is less frequent in many developing countries. Input-Output data is not available for all countries. While these datasets are constructed following common principles and classifications, data availability and differences in the aggregation level and collection frequency can limit comparability and the research questions explored, and guide further modelling choices. For example, in many developing countries, expenditure datasets do not contain income data (relevant for the definition of the welfare concept) and report products at inconsistent levels of aggregation (relevant for simulating direct emissions). Input-output tables can be severely outdated, and datasets may differ in the available environmental extensions (relevant for simulating indirect emissions). We return to these issues in Sections \ref{coverage} and \ref{inc_concept}. The disproportionate representation of upper-middle and high-income countries may also be explained by the high cost of developing these models and greater policy interest in carbon taxation in these countries.

The number of countries analysed can also be guided by practical considerations and the research question. Including as many countries as possible may be preferable when the purpose is to produce generalisable insights, but may also make it difficult to understand what drives results within each country and to validate results. Analysing a few countries only can have the advantage that it allows researchers to investigate the drivers of distributional impacts within countries and to compare drivers across countries.

\subsection{Tax coverage and design}\label{coverage}
The importance of tax coverage and differential tax rates on energy goods for the distributional impact of energy taxation has been subject to much discussion \citep{harding2014taxing, flues2015distributional}.  

\cite{flues2015distributional} and \cite{barker1998equity} show that taxing heating fuel and electricity is more regressive than taxing motor fuels. Studying 11 EU countries, \cite{vandyck2014distributional} show that a carbon price on transport fuels is regressive in most countries, but progressive in Estonia. \cite{datta2010incidence} find that a carbon tax is progressive in India, with the only regressive component relating to kerosene. \cite{jacobs2022distributional} show that taxing diesel is more progressive than taxing gasoline. 

In the US, \cite{grainger2010pays} show that taxing all goods is less regressive than taxing energy goods only. In their meta-analysis, \cite{ohlendorf2021distributional} confirm that including indirect emissions increases the likelihood of finding progressive outcomes. \cite{steckel2021distributional} show that the distributional impact of a globally harmonized carbon price, national carbon prices, and sectoral carbon prices of the power and transport sector differ strongly across designs and countries.

In a recent study on Belgium, \cite{bursens2026bridging} simulate various designs of a carbon tax on energy products, including differentiated rates by fuel, a progressive rate structure, and flat rates with a GHG emission allowance. They highlight difficulties in offsetting carbon tax regressivity through design only, and point towards the best trade-off between regressivity and effectiveness when rates are differentiated across products.

The Greenhouse Gas (GHG) emissions covered by the tax also influence its distributional impact, with $CO_2$ emissions resulting primarily from energy use, and $CH_4$ and $N_2O$ emissions from agricultural production \citep{sager2019income, mardones2020economic, hynes2009spatial}.  \cite{kerkhof2008taxation} find less regressive impacts when GHG emissions other than $CO_2$ are covered. The inclusion of Methane emissions ($CH_4$) and Nitrogen dioxide ($N_2O$) can also increase carbon tax regressivity through increased food prices \citep{renner2018poverty, vogt2019cash}.

\subsubsection{Implementation: Energy Data Granularity and Input–Output Choices}

\textit{Direct emissions}

Modelling direct emissions from fuel combustion requires information on energy expenditures and prices (or quantities) and carbon intensity factors. Frequently, expenditure datasets only report expenditure, but not quantities. As carbon taxes are levied in relation to quantities, computing direct emissions requires price information to estimate quantities. Highly volatile energy prices mean that the carbon emissions associated with 1€ of energy expenditure can vary substantially over time, potentially leading to divergent estimates when the same expenditure dataset is used. However, energy price data or price changes per commodity are rarely reported (exceptions include \cite{wier2005co2} and \cite{kerkhof2008taxation}). This complicates comparisons of direct emission factors (per monetary unit) across studies. This is further complicated by the fact that some popular databases (e.g. GTAP) are not publicly available.

An additional challenge relates to the level of aggregation of energy commodities in expenditure surveys and EE-IO datasets. Sometimes, emission factors or carbon tax payments are reported for aggregate groups, such as domestic or transportation/motor fuels. Heating-related emissions crucially depend on the fuel mix. Authors rarely report the extent to which energy commodities are differentiated, and differences in the carbon intensity of the energy mix consumed are rarely shown explicitly (except in \cite{verde2009distributional, linden2024manyfaces, jaccard2021energy}). 

The treatment of biomass and solid fuels can also significantly impact estimates, with some studies exempting them\footnote{Exemptions are commonly motivated by difficulties in taxing biomass in practice or their carbon neutrality in line with IPCC assumptions.} \citep{missbach2024cash, dorband2019poverty} while others include them \citep{steckel2021distributional}. Biomass such as firewood, dung, turf or peat has high carbon intensity factors per Kwh, is often cheaper than fossil fuels, and is primarily consumed by low-income households \citep{dorband2019poverty}. They can therefore be an important determinant of households' carbon footprint. \\

\textit{Indirect emissions}

Indirect emissions are commonly estimated using the IO methodology \citep{leontief51, leontief1974sructure}. IO models provide information on the input requirements of one industry (in one country) from all other industries (in the country and all other countries), and trace (monetary) flows across industries (and countries). To model indirect emissions, IO models are complemented with environmental extensions capturing the environmental impact per unit of industry output \citep{feindt2021understanding, steckel2021distributional, dorband2019poverty}. \cite{Kitzes2013} introduce environmentally extended IO models (EE-IO) and \cite{minx2009input} provide an overview of applications. \cite{huang2019systematic} review methods to estimate sectoral emissions, including EE-IO. 

Broadly, we can differentiate between five approaches to modelling indirect emissions. The first approach is to ignore indirect emissions and to only consider direct emissions (e.g. \cite{grainger2010pays}). The second approach is to use a single-region IO, either exempting imported goods from the tax or assuming no significant difference between imported and domestically produced goods. The third approach is to use Multi-regional IO models (MRIO) \citep{miller2009input}, mapping trade flows across industries and countries, accounting for cross-country differences in the production technology and carbon intensity \citep{sager2023global}. The fourth approach is to use emission estimates provided directly by national statistical institutes \citep{levay2021association}. The fifth approach is to rely on estimates from the literature \citep{eisenmann2020distributional}.

The use of MRIO models is becoming increasingly common and multiple MRIO databases provide environmental extensions (emissions per sector). The most popular databases include GTAP, WIOD, EXIOBASE, and EORA \citep{wiedmann2011quo}. The database used can affect estimates \citep{steen2014effects, tukker2013global} because of differences in the sectoral level of aggregation, in the environmental extension available ($CO_2$ or other GHG emissions, water, and land use), and methodological differences in their construction \cite{hamilton1994simulating}. 

The level of industry aggregation impacts IO coefficients, multipliers \citep{kymn1990aggregation}, and distributional estimates. The level of energy sector aggregation is particularly relevant \citep{steen2014effects, lenzen2011aggregation}\footnote{Differences in the level of aggregation between IO data and environmental data may force researchers to aggregate environmental data to the IO sector level or to disaggregate the IO sector to resemble the environmental data \citep{lenzen2011aggregation}. \cite{lenzen2011aggregation} suggests that disaggregation of IO data, even if based on limited data, is preferable over the aggregation of environmental data in determining IO multipliers.}. For example, GTAP differentiates between electricity and gas supply, while the sectors are grouped in WIOD. Overall, WIOD and GTAP previous releases produce similar results in most regions \citep{owen2014structural}, but estimates may differ significantly across some countries \citep{arto2014comparing}. 

Differences in emission datasets also cause divergent estimates \citep{peters2012synthesis}. Differences arise from the allocation principle (territorial principle vs residence principle), the level of aggregation in the emission vector, or the source of the emission estimates. GTAP emission vectors capture $CO_2$ emissions derived from IEA energy data and focus on emissions from fossil fuels (GTAP 7), whereas WIOD uses data from NAMEA and also includes process-based and fugitive emissions \citep{owen2014structural, arto2014comparing, corsatea2019world}. Lastly, the reference year of the datasets also impacts estimates. 

In selecting the appropriate IO model (Single region or Multi-region) and database (WIOD, GTAP, EXIOBASE, EORA), modelers should study the benefits and drawbacks of each approach and database, and make motivated choices based on the application of interest \citep{kymn1990aggregation, steen2014effects, owen2014structural, arto2014comparing}.

\subsubsection{Behaviour}

Carbon pricing increases the cost of pollution to incentivize households and firms to reduce their emissions \citep{fullerton2007general}. Accounting for behavioural responses is central to estimating the emissions, welfare, and distributional impacts of carbon pricing, yet these responses are heterogeneous and difficult to model empirically. Four types of behaviour are particularly relevant for the functioning of a carbon tax. 

First, households reduce consumption of carbon-intensive goods in response to higher prices, reflecting both substitution and real-income losses (changes along the intensive margin). Second, households invest in new technologies, like electric vehicles or heat pumps, or change practices, like using public transportation, to avoid higher fuel costs (changes along the extensive margin). Third, when carbon tax revenues are returned to households, they allocate additional income across products\footnote{The way households allocate additional income (expenditure) across products is captured by income (expenditure) elasticities. For a review of this literature and a discussion of methodological considerations, we refer to  \cite{pottier2022expenditure} and \cite{levay2023income}.}. 

Fourth, firms respond to higher energy prices by substituting away from carbon-intensive inputs (firm-level behavioural responses). In general equilibrium (GE), these adjustments interact with endogeneous changes in output, relative prices, wages, and rents, as goods and factor markets clear simultaneously and income and demand feedbacks propagate economy-wide.

Here, we focus on the intensive margin and GE effects, distinguishing between three approaches: focusing on immediate impacts (ignoring firm and household behaviour), modelling household behaviour (in a partial equilibrium setting), and modelling household and firm behaviour (in a GE setting). We discuss models of the extensive margin and behavioural responses to revenue recycling in section \ref{FurtherResearch}.

Many microsimulation studies are arithmetic (static) and assess the immediate distributional impacts of carbon taxation, abstracting from firm and household behaviour. These studies are often viewed as providing an upper-bound estimate of the carbon tax burden\footnote{Examples include \cite{hynes2009spatial, grainger2010pays, christis2019detailed, saelim2019carbon, wier2003environmental, sager2019income, callan2009distributional, jiang2014distributional, kerkhof2008taxation, verde2009distributional, hassett2009incidence, mathur2019rethinking, eisenmann2020distributional, winter2023carbon, brenner2007chinese, missbach2024cash, dorband2019poverty, steckel2021distributional, vogt2019cash, rub2024inequality, jaccard2021energy}}. Some studies include intensive margin behavioural responses by households in a partial equilibrium setting but abstracting from economy-wide feedbacks\footnote{Examples include \cite{creedy2006carbon, semet2024coordinating, tiezzi2005welfare, reanos2022distributional, rosas2017distributional, saelim2019carbon, renner2018household, douenne2020vertical, labandeira2006residential, berry2019distributional, yusuf2015distributional, linden2024manyfaces}}. Fewer distributional analyses account for endogenous price adjustments, factor returns, and producer responses by relying on Computational General Equilibrium (CGE) models to jointly capture the response of consumers and producers\footnote{Examples include \cite{Antosiewicz22, fragkos2021equity, chepeliev2021distributional, vandyck2021climate, barker1998equity, campagnolo2022distributional, yusuf2015distributional, mardones2020economic, hamilton1994simulating, williams2014initial, zhang2019effects, rausch2011distributional, rausch2016household, goulder2019impacts, dissou2014can}}. 

In partial equilibrium, the distributional impact of household behaviour along the intensive margin depends on multiple factors; the price sensitivity of demand for carbon-intensive goods and energy services, systematic differences in these elasticities across the income distribution, and the budget shares devoted to these expenditures. Together, these mechanisms shape how carbon price increases translate into changes in consumption behaviour and thus differential burdens.

Incorporating household behaviour along the intensive margin often reduces carbon tax regressivity \citep{ohlendorf2021distributional} because poorer households are budget-constrained and likely more responsive to price changes. \cite{west2017estimates} find stronger responses among low-income consumers, while \cite{zhang2015energy} and \cite{campagnolo2022distributional} find stronger responses among high-income households. Other household characteristics also shape household behavioural responses. \cite{lawley2018refining} find weaker responses of rural households to carbon-tax-induced gasoline price changes in British Columbia. \cite{eisner2021distributional} show that age, region, and dwelling age are important for the response to heating fuel price changes, while household composition is important for electricity consumption, and location for motor fuels. Overall, the impact of intensive margin household behaviour for carbon tax regressivity is likely limited \citep{renner2018household, labandeira1999combining}. More important behaviour change likely relates to the extensive margin. Such changes remain underexplored in the literature.

The distributional impact of GE effects depends on the distribution of production factors, capital, and labour \citep{hamilton1994simulating}, and on how carbon taxation changes the returns to these factors. \cite{metcalf2023five} argues that carbon taxes are progressive in source-side effects (changes in capital and labour market returns) and often regressive in use-side effects (changes in consumer price). When changes in relative factor prices favour (low-skill) labour and reduce the returns to capital \citep{williams2014initial}, they lead to more, or overall, progressive outcomes as found in \cite{mathur2014distributional, rausch2011distributional, goulder2019impacts}. Others find the opposite \citep{wu2022distributional, landis2019cost}, or mixed results depending on the region \citep{williams2014initial}. In their meta-analysis, \cite{ohlendorf2021distributional} do not find a significant impact of GE effects on carbon tax regressivity. 

Static models provide an approximation of the day-after effects on households' living costs, keeping all else equal. Partial equilibrium studies complement this insight with households' ability to avoid carbon-intensive goods. GE studies provide the most complete analysis of the effects of a carbon tax as they account for economy-wide adjustments. They however ignore the experience of households throughout the adjustment period.

\subsubsection{Implementation: Price elasticity estimation and Micro–Macro Linkages}

\textit{Demand-side: Household Behaviour along the intensive margin}

Household behaviour is commonly modelled by estimating expenditure systems on household expenditure data or by 'borrowing' estimates from elsewhere\footnote{Useful sources of elasticity estimates include \cite{labandeira2017meta}'s meta-analysis on price elasticity of energy demand and \cite{temursho2024consumer}'s estimates for the EU.}. Popular expenditure systems include the Linear Expenditure System (LES) \citep{stone1954linear}, the (quadratic) Almost Ideal Demand System (QAIDS) \citep{deaton1980economics, banks1997quadratic}, and the Exact Affine Stone Index (EASI) \citep{lewbel2009tricks}. These expenditure systems estimate own-price, cross-price, and income elasticities by exploiting the relationship between expenditure shares and income (Engel curves), price fluctuations, and resulting changes in expenditures. 

The choice of expenditure system can affect estimates and conclusions. \cite{reanos2022measuring} find more regressive results using an EASI compared to a QUAIDS model and suggest that QUAIDS underestimates the distributional effect of carbon taxation. The choice of expenditure system can be practical. Most expenditure systems require information on prices and expenditures. In the absence of price data, the LES can be estimated by imposing theoretical restrictions and using cross-sectional expenditure data and the marginal utility of income with respect to income (the ‘money flexibility’ or Frisch parameter) \citep{frisch1959complete, cornwell1998measuring, temursho2024consumer}. A common approach to estimate more flexible AIDS, QUAIDS, or EASI models is to pool multiple years from cross-sectional household expenditure surveys and to source Consumer Price Indices (CPI) per expenditure group for these years \citep{tovar2023benefits, renner2018household, semet2024coordinating}. 


Table \ref{elast} provides a non-exhaustive list of average own-price elasticity estimates from the literature. In all studies, except those conducted in France, own-price elasticities of heating fuels are smaller than those on motor fuels, though this varies along the income distribution and other household characteristics, such as location \citep{creedy2006carbon, ravigne2022fair, douenne2020french, moz2021winners, eisner2021distributional}. 

The estimates in Table \ref{elast} can be compared to ex-post estimates \citep{lin2011effect, andersson2019carbon, davis2011estimating}. Studying Sweden's carbon tax and fuel VAT tax, \cite{andersson2019carbon} estimates an 11\% reduction in transport sector $CO_2$ emissions and a carbon tax elasticity of -1.57; three times higher than the own-price elasticity. This confirms findings by \cite{rivers2015salience} and \cite{li2014gasoline} who show that consumers respond more strongly to price changes due to taxes than to exogenous price shocks. Structural estimates (from expenditure systems) may therefore significantly underestimate the environmental benefits of carbon taxation. 


\begin{center}
\adjustbox{max width=\textwidth}{
\begin{threeparttable}
\caption{Non-exhaustive overview of own-price elasticities estimates in the Literature}
\label{elast}
\large
\begin{tabular}{ l l l l l l l l}
  \hline
         Paper & Country & Energy & Electricity & Heating fuels & Motor fuels & Demand system \\ \hline\hline
        \cite{reanos2022measuring} & Ireland & 	-0.47 &   &  & -0.49$^1$ & EASI \\ 
        \cite{tovar2023benefits} & Ireland &  & -0.46  & -0.46 & -0.44 & EASI \\ 
        \cite{tiezzi2005welfare} & Italy &  &  & -1.06 &  -1.28 & AIDS \\ 
        \cite{labandeira2006residential} & Spain &  & -0.78  & -0.05$^2$  & -0.06 & QUAIDS \\ 
        \cite{rosas2017distributional} & Mexico &  & -0.64  & -0.35 & -0.73   & AIDS \\
       \cite{renner2018household} & Mexico &  &  -1.49 &  -0.69$^3$ &  -1.03  & QUAIDS  \\
        \cite{berry2019distributional} & France &  &   & -0.35 & -0.18  & Engel curve model \\
        \cite{douenne2020vertical} & France &  &   & -0.54$^4$ & -0.28$^4$  & QUAIDS \\
         \cite{semet2024coordinating} & France &  &  & -0.20$^5$ & -0.23$^5$  & QUAIDS \\
        \cite{saelim2019carbon} & Thailand &  & -0.53  &  & -0.60  & QUAIDS \\
     \cite{brannlund2004carbon} & Sweden &  &  & -1.81 & -1.18  & QUAIDS \\
     \cite{moz2021winners} & Brazil & -0.93 &  &  & -0.58  & Censored QUAIDS \\
   
    \cite{linden2024manyfaces}& six EU$^6$  &  & -0.26 & -0.38   & -0.50  & LES \\\hline
    \cite{labandeira2017meta} & Meta-analysis & -0.22 & -0.13  & -0.18$^7$ & -0.29$^7 $ & Literature review \\ 
    \cite{espey1998gasoline} & Meta-analysis &   &  &  &  -1.37 to 0 & Literature review \\ 
    \cite{espey2004turning} & Meta-analysis &   & -0.35 &  &   & Literature review \\ 

    \cite{brons2008meta} & Meta-analysis &  &  &   &  -1.36 to 0.37 & Literature review \\ \hline
    \cite{andersson2019carbon} & Sweden& -1.57$^8$ &  &  &  & Ex-post \\ 
    \cite{rubin2024quantifying} & California & &  &  -0.19 to -0.15$^5$ &  & Ex-post \\ 
    \cite{maddala1997estimation} & USA &  &  & -.18 to -.09$^5$ &    & Ex-post \\ 
     \cite{davis2011estimating} & USA &  &  & &  -0.46 to -0.1$^9$  & Ex-post \\ 
    \cite{rehdanz2007determinants} & Germany &  &  & -0.63 to -0.44$^5$ &    & Ex-post \\ 
    \cite{baranzini2013elasticities} & Switzerland &  &  &  & -0.37 to -0.09$^9$  & Ex-post \\ \hline\hline

    \end{tabular}
	\begin{tablenotes}
      \item Note: Values shown in this table represent short-term elasticities. Where possible, we show elasticities for population average values, evaluated at the mean value of budget shares. Where both compensated and uncompensated elasticities are available, we show uncompensated elasticities. Where estimates were available for the total sample of households and a sub-sample of households that consume the good, we show the value for the sub-sample of households that consume the good. (1) Represent the values for the first quartile of total expenditure. The values for the fourth quartile is -0.538 and -0.742 for energy and motor fuels (transport), respectively. (2) Natural Gas, the value for LPG is -0.249. (3) Natural Gas. (4) Represent elasticities for households of the 5th decile in medium-sized cities, (5) Home fuel represents Domestic Energy, including electricity. Motor fuels represent transportation, including public transportation and air travel. (6) Pooled across Hungary (-0.36, -0.37, -0.74), Lithuania (-0.19, -0.37, -0.35), Portugal (-0.23, -0.49, -0.46), Ireland (-0.12, -0.29, -0.35), Finland (-0.28, -0.28, -0.40), and Luxembourg (-0.06, -0.26, -0.32), population weights applied. Country estimates for 10 household types. Numbers behind countries refer to the country average value per commodity (7) Natural Gas and Gasoline, respectively; the values for heating oil and Diesel are -0.29 and -.15. (8) Carbon tax elasticity of demand.  (9) Gasoline. (10) Short-term mean estimate, Median -0.28, ranging from -2.01 to -0.004

    \end{tablenotes}
\end{threeparttable}}
\end{center}

\textit{Supply-side: General equilibrium effects}

Similar to IO models, Computational GE (CGE) models use IO tables and social accounting matrices as input data, but allow for more flexibility in modelling agents' behaviour through the specification of functional forms and the modelling of factor substitution. To assess the distributional impacts of carbon pricing in CGE models, the household sector (frequently represented by a single agent) needs to be disaggregated. \cite{van2015methods}, \cite{bourguignon2008impact}, \cite{bourguignon2008distributional} and \cite{cockburn2014macro} review methods to include household heterogeneity in CGE models. 

There are three common approaches to integrating household heterogeneity into macro-economic models; imposing heterogeneity through assumed or empirically observed income distributions within an otherwise representative-agent framework, specifying multiple representative households differentiated by, for example, income, skill, or sector, and linking macroeconomic models with household-level microsimulation models. We focus on the latter two. 

A common approach is to integrate multiple household types characterized by differences in income, location, employment, or expenditure patterns \citep{fragkos2021equity, EKINS20112472}. The level of heterogeneity considered affects distributional impact estimates by influencing how returns to production factors respond to a carbon tax \cite{rausch2016household}. 


It may therefore be preferable to prioritize household heterogeneity by integrating microsimulation and CGE models \citep{rausch2011distributional, mardones2024contribution}. Directly integrating a representative sample of households into a macroeconomic model is theoretically the soundest \citep{bourguignon2008impact}, with likely the most ambitious effort made by \cite{rausch2011distributional}, who integrated 15,000 households into a macroeconomic model to compute feedback effects from households to firms. 

Alternatively, micro-and macroeconomic models can be linked through a "soft link" \citep{labandeira2009integrated, Antosiewicz22}, whereby model outputs are exchanged (sometimes iteratively) between the micro- and macro-model. The economic impact of a carbon tax is first estimated using a CGE model, and the micro-data is reweighted to match the aggregates produced in the macro model \citep{chepeliev2021distributional, vandyck2021climate, ravigne2022fair}. \cite{dissou2014can} compute the impact of a carbon tax on commodity and factor prices using CGE models and pass changes in commodity prices and factor returns through to household-level datasets. 

Models sometimes have a bottom-up (also sometimes called 'backward' or 'upward') linkage from the microsimulation model to the macroeconomic model. These models update the projected aggregate of key variables using a factor reflecting changes in the aggregate of variables in the micro-dataset. In practice, this is often implemented through iterative reweighting or scaling factors that ensure consistency between micro-level distributions and macroeconomic projections. For example, \cite{ravigne2022fair} use an iterative bi-directional five-step “soft-link” macro–micro framework to project the distributional impact of carbon taxation up until 2035. They pass aggregate growth factors for income components, taxes, and relative prices down from IMACLIM-3ME (the macro-economic model) to MATISSE (the microsimulation model). The population of households is reweighted to reproduce aggregate IMACLIM-3ME evolutions of total labour income, capital income, unemployment benefits, other social benefits, remittances, and aggregate direct taxes (including income taxes), as well as demographic dynamics taken from the French statistical office’s population projections. A central novelty of \cite{ravigne2022fair}'s model is the explicit incorporation of technology diffusion. EV adoption and dwelling renovations are allocated to households in line with aggregated projections, updating household energy demand, electricity use, investment and loan repayment costs, and savings. Next, household behaviour is simulated in MATISSE (the microsimulation model), and the resulting energy consumption, total consumption, and savings are aggregated are fed back into IMACLIM-3ME. The macro–micro loop is iterated until consistency between aggregate macroeconomic outcomes and micro-level behaviour is achieved. \cite{ravigne2022fair} do not explicitly model labour market transitions or sectoral employment reallocation, and assume technology diffusion trajectories rather than modelling them.

In contrast to aggregation-based or iterative macro–micro frameworks, \cite{Antosiewicz22} rely on structural integration rather than aggregation-based feedbacks. Focusing on job flows, \cite{Antosiewicz22} embedded a (hard-linked) microsimulation within a dynamic stochastic general equilibrium (DSGE) model featuring a search and matching labour market model, allowing labour market frictions and employment transitions to be modelled consistently at both micro and macro levels. In this set-up, employment transitions and income dynamics are governed directly by macro-consistent transition equations. 

\cite{rausch2011distributional} go further and pursue full structural integration by representing heterogeneous households explicitly as agents within a computable general equilibrium framework. In this set-up, distributional outcomes arise endogenously from the macro-economic model's equilibrium solution rather than from aggregation or feedback between separate macro and micro models. While \cite{rausch2011distributional} ensure macro–micro consistency by fully integrating heterogeneous households into the CGE equilibrium, \cite{Antosiewicz22} ensure consistency by projecting DSGE-generated sectoral shocks onto household microdata without macro feedback. As noted by \cite{Antosiewicz22}, solving a dynamic DSGE model with the same level of household heterogeneity as in \cite{rausch2011distributional} would be computationally prohibitive, which helps explain why fully integrated CGE–micro approaches remain rare and largely static.

\subsection{Revenue recycling}\label{revenue}
The distributional effects of revenue use outweigh those of carbon taxation itself \citep{fremstad2019impact, klenert2018making}, making revenue recycling the key driver of carbon taxation's overall distributional outcomes. For example, returning tax revenues to households as per capita transfers leads to progressive impacts and can reduce poverty \citep{klenert2018making, malerba2021mitigating}, while using revenues for more complex reforms, such as tax cuts or subsidies, can be progressive or regressive depending on their design \citep{brannlund2004carbon, missbach2023assessing}. 

Revenue recycling was first discussed in the 1990s under the double dividend hypothesis, by which a carbon tax combined with reductions in distortionary taxes could simultaneously improve economic and environmental conditions \citep{pearce1991role, goulder1995environmental}. The distributional impacts of such 'environmental tax reforms' have been assessed in various countries \citep{goulder1995effects, barker1998equity}. The distributional impact of environmental tax reforms depends on what taxes are reduced and for whom. Labour tax cuts can be distribution-neutral to mildly regressive \citep{barker1998equity, EKINS20112472}, while recycling via corporate or capital tax cuts is more likely to be regressive \citep{caron2018distributional, williams2015initial, mathur2014distributional}.


The focus in the recent literature has shifted towards direct compensation for increased energy prices and living costs (Figure \ref{FIG:revrecyOptions} of the supplementary materials shows the frequency of schemes modelled.). The most widely studied scheme is a per capita lump-sum transfer \citep{Antosiewicz22, grainger2010pays} (\ref{FIG:revrecyOptions}). Energy poverty and geographically targeted transfers are also analysed more often \citep{douenne2020vertical, berry2019distributional, reanos2022measuring}. Other simulated schemes include indirect tax cuts \citep{symons1994carbon, brannlund2004carbon, missbach2023assessing}, public transport subsidies \citep{brannlund2004carbon, buchs21, missbach2023assessing}, infrastructure financing \citep{dorband2022double}, wage subsidies \citep{Antosiewicz22}, energy-efficiency subsidies \citep{giraudet2021policies, bourgeois2021lump}, electricity and transport fuel subsidies \citep{missbach2023assessing}, and social benefits increases \citep{callan2009distributional, verde2009distributional, vandyck2021climate}.


\subsubsection{Implementation: Data Linkages for Revenue Recycling}

Universal and targeted transfers can be implemented using expenditure survey data only. More complex revenue recycling schemes use information not typically included in these datasets (income sources, including wages, capital incomes, benefits, and tax liabilities) and require matching across datasets. 


Authors have used various approaches to integrate relevant but missing information into expenditure datasets. \cite{callan2009distributional} and \cite{verde2009distributional} compute carbon tax payments and additional benefit payments using separate datasets and compare average carbon tax and benefit payments across disposable income deciles. Using a nested regression-based approach, \cite{immervoll2023pays} impute expenditure patterns from EU-HBS into the EU Survey on Income and Living Conditions (SILC) and adjusts benefits and tax payments computed in the tax-benefit simulation model EUROMOD. \cite{reanos2021fuel} compute a Stone index using HBS and allocate the index to SILC data based on income deciles and household types. \cite{hynes2009spatial} directly match the Census of Agriculture to a National Farm Survey using simulated annealing. \cite{douenne2020vertical} use a non-parametric Nearest Neighbour (NND) hot-deck procedure to match the French transport survey to its consumer expenditure survey. \cite{vandyck2021climate} use the specialized indirect tax tool (ITT), an extension of EUROMOD, to simulate various revenue recycling options. The ITT mixes imputation and matching techniques (predictive mean matching) to introduce expenditure patterns from the EU-HBS into the EU-SILC \citep{akouguz2020new}. Recently, the EUROMOD team has developed the Green EUROMOD, which largely follows the approach described in \cite{akouguz2020new}, complementing the EUROMOD datasets with expenditures (used by the Consumption Tax tool), as well as linking the dataset with the EXIOBASE MRIO (see \cite{bursens2026bridging} for an application.). An alternative is to generate synthetic datasets mirroring the population by combining information from different data sources and using algorithms such as Iterative Proportional Fitting and Random Forests \citep{tobben2023unequal}. 

The data used by GE models commonly contains economy-wide data that explicitly represent labour markets, including labour income, wages, and labour taxes. This structure allows analysts to simulate labour tax reductions directly within the macroeconomic framework, without relying on micro-level imputation or matching procedures (for example, GEM-E3-FIT model \citep{fragkos2021equity}). 


This review only included a few papers investigating the distributional impact of green subsidies in microsimulation frameworks, as they require specialized datasets. The French Phebus dataset contains information on the energy performance of the housing stock alongside data on occupants' characteristics, their assets, and energy use \citep{bourgeois2021lump, giraudet2021policies, berry2019distributional}. Mobility in  Germany and the German Mobility Panel in Germany provide information on the everyday mobility of Germany’s residential population \citep{jacobs2022distributional}. While not strictly related to carbon taxation or microsimulation modelling, \cite{borenstein2016distributional} use tax return data to examine the socioeconomic characteristics of US Clean Energy Tax Credits and, more recently, US Tax Credits for Heat Pumps, Solar Panels, and Electric Vehicles \cite{borenstein2025distributional}. 

The choice of revenue recycling scheme also affects other modelling choices. For example, when modelling direct compensation, it is intuitive to use disposable or current income as a welfare measure and ranking concept as transfers commonly form one component of income (see section \ref{inc_concept}). Using expenditure requires assumptions on how and how much of the transfer is spent. Estimated demand systems or sourcing elasticities from the literature can help refine this assumption.

\subsection{Welfare}\label{inc_concept}
Distributional impacts describe the distribution of costs and benefits across population groups. Costs and benefits are typically computed relative to a welfare measure, often current income or expenditure, and compared across income groups. Both the variable used to measure welfare (the welfare concept) and the variable used to construct groups (the ranking concept) affect distributional impact estimates \citep{shei2024distributional}. \\

\textit{Choice of Welfare Concept: Income vs. Expenditure}

The most widely used welfare concepts in distributional analysis of carbon taxation are current (annual) income and expenditure. While both are often used to approximate economic welfare, the choice between them affects estimated distributional impacts and their interpretation, reflecting differences between short-run income and longer-run consumption smoothing \citep{poterba1991tax}. Using annual income rather than expenditure as a welfare concept often leads to more regressive results \citep{davies1984some, poterba1991tax, grainger2010pays, ohlendorf2021distributional}. A common critique of using annual income as a welfare concept is that it overstates the carbon tax burden of households experiencing temporarily low incomes due to job loss, the life cycle (students or the retired), or seasonal variation in incomes \citep{poterba1991tax, hassett2009incidence, pizer2019distributional}. Therefore, some authors use expenditure as a welfare concept to approximate lifetime income. Theoretically, this approach is underpinned by \cite{friedman1957theory}'s permanent income hypothesis by which households smooth their consumption over their lifetime, assuming perfect information on lifetime income streams and credit market access. \cite{rausch2011distributional} find no evidence that using annual income biases the results towards higher regressivity, while \cite{ohlendorf2021distributional} find that a significantly higher likelihood of finding progressive results when lifetime income proxies are used.




\textit{Choice of Ranking Concept: Household Ordering and Distributional Outcomes}
\\
The ranking concept affects households' position along the income distribution \citep{lambert2001distribution} and consequently the variation in carbon tax burdens across income groups. \cite{shei2024distributional} and \cite{hassett2009incidence} demonstrate that disposable income-based rankings amplify the regressive effects of a carbon tax. \cite{shei2024distributional} suggest that using expenditure as welfare and ranking concept is superior to using disposable income because the use of expenditure results in more stable distributional outcomes, though they caution against using one-size-fits-all approaches. Recently, \cite{buchs2024emission} study carbon emissions along the wealth distribution by imputing carbon emissions into datasets containing individual-level wealth data. Their study shows that the emission inequality changes across the expenditures, income, wealth, and the joint income and wealth distributions. \cite{buchs2024emission} find that per capita emissions are more unequally distributed over expenditure than income distribution, and most unequally distributed over the joint wealth and income distribution. \\

\textit{Interpretation of Carbon Tax Burdens Across Time Horizons}
\\
Disposable income and expenditure are often treated as interchangeable, but their use affects the interpretation of a carbon tax burden and implies different time horizons. Annual or disposable income better approximates short-term ability-to-pay \citep{chernick1997pays}, households' ability to absorb the increase in living costs without dis-savings or borrowing, and has been referred to as the effort rate \citep{semet2024coordinating}. Carbon tax burdens computed using expenditure reflect the carbon intensity of the consumption basket and the households' exposure to carbon pricing. The difference between disposable income and expenditure is given by the savings rates (flow of savings), with negative or low rates at lower incomes \citep{dynan2004rich}. \cite{linden2024manyfaces} show that differences in savings rates along the income distribution are an important driver of regressivity in many countries. Accounting for savings rates is therefore likely important for capturing liquidity constraints and estimating short-run carbon tax burdens, particularly for population groups that have accumulated lower savings stocks and have poorer access to credit markets (e.g. the young and those with low earnings potential). In the medium to long term, using current income risks overstates the carbon tax burden of households with temporarily depressed incomes as economic agents smooth their consumption, so that expenditure reflects permanent income and long-run welfare.

Figure \ref{FIG:combined_CT_Decile_HU} illustrates how welfare concepts and ranking variables impact distributional impact estimates for Hungary and proposes an interpretation for each combination of welfare and ranking concept. The Panel (a) shows tax payments as the share of disposable income along the income distribution and can be taken as the immediate financial stress experienced by households across income levels. Panel (b) shows tax payments as the share of expenditure along the income distribution and can be taken as the carbon intensity of the consumption basket along the income distribution, with the difference between (a) and (b) given by the savings rate. Panel (c) shows the carbon tax as a share of income along the expenditure distribution and captures immediate financial stress along the permanent income distribution. Ranking households by expenditure may help address concerns over temporarily low incomes, as households with high (low) permanent incomes are likely located at the top (bottom) of the expenditure distribution, but simultaneously captures the immediate financial stress experienced by households with low current incomes. Lastly, Panel (d) shows the tax payments as a share of expenditure along the expenditure distribution and can be interpreted as carbon economies of scale, i.e. the relative carbon intensity of necessities relative to luxury goods. 


\begin{figure}[ht]
    \caption{Comparison of carbon tax burdens by welfare concept and ranking variable for Hungary.}
	\label{FIG:combined_CT_Decile_HU}
	\centering
		
		\subfloat[Income share across income deciles. Immediate financial stress across current income groups\label{fig:f3}]{
		\copyrightbox[b]{\includegraphics[width=0.44\textwidth]{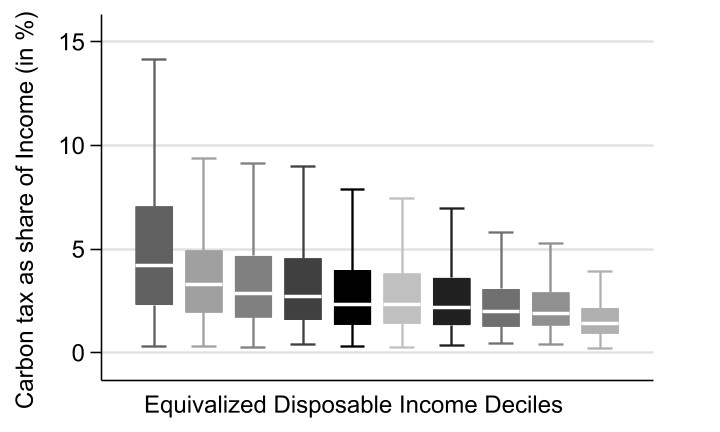}}%
		{}
		}%
		\qquad
		\subfloat[Exp. share across income deciles. Carbon intensity of consumption across current income groups\label{fig:f4}]{
		\copyrightbox[b]{\includegraphics[width=0.44\textwidth]{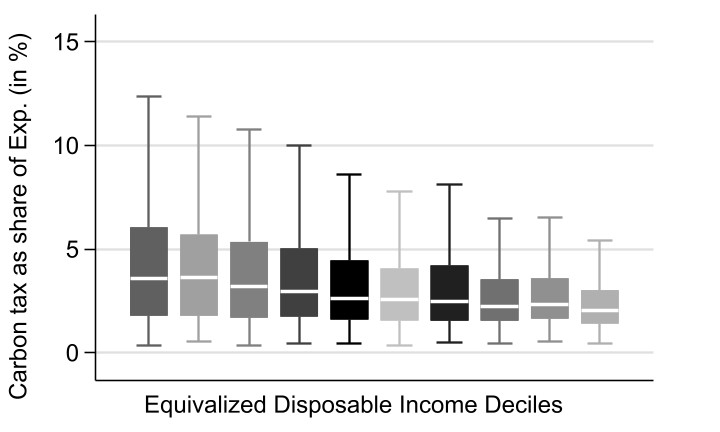}}%
		{}
		}%
		
		\subfloat[Income share across exp. deciles. Immediate financial stress across lifetime income groups\label{fig:f5}]{
		\copyrightbox[b]{\includegraphics[width=0.44\textwidth]{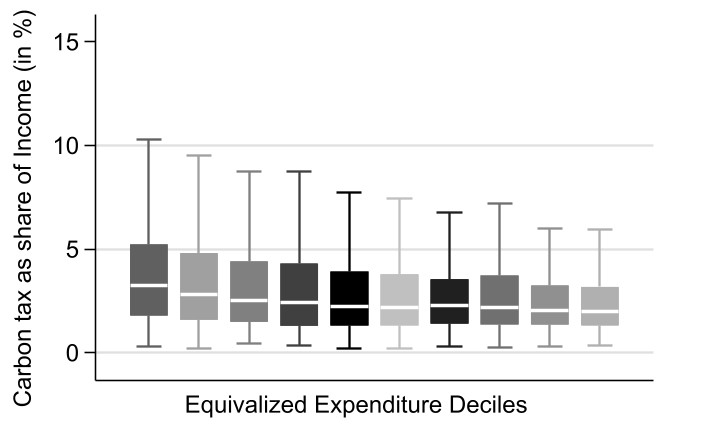}}%
		{}
		}%
		\qquad
		\subfloat[Exp. share across exp. deciles. Carbon intensity of consumption across life-time income groups\label{fig:f6}]{
		\copyrightbox[b]{\includegraphics[width=0.44\textwidth]{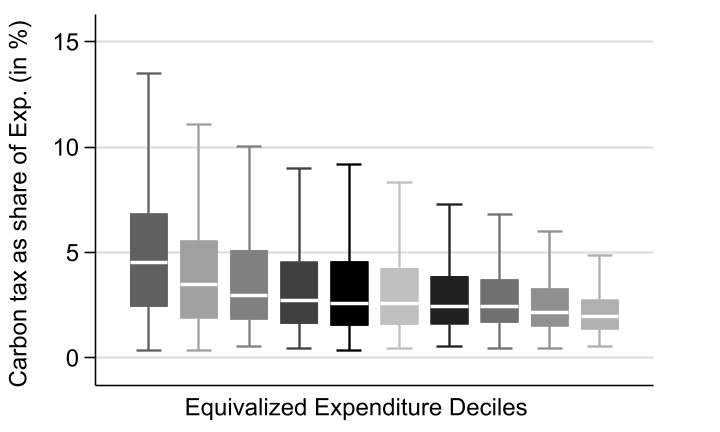}}%
		{}
		}%

		\caption*{\footnotesize Carbon tax payments (€30/tCO2) as a share of household disposable income and expenditure (welfare concept) along the equivalized disposable income and equivalized expenditure deciles (ranking variable) in Hungary using the 2015 EU-HBS. Authors' own calculations following the methodology described in \cite{odonoghue2024distributional}.}
	\label{FIG:combined_CT_Decile_HU}
\end{figure}

\FloatBarrier



\subsubsection{Implementation: Income Fluctuations and Alternative Welfare Metrics}

\textit{Dealing with temporarily low incomes}

The discussion on the appropriate welfare measure primarily centers around temporarily low incomes \citep{poterba1991tax} and unreliable incomes in survey data \citep{blundell1998consumption}. While the most popular approach is to approximate lifetime incomes with expenditure, alternatives exist.

\cite{rausch2011distributional} restricts the sample to households where the head of household is between 40-60 years old, excluding pensioners and students, who often have low temporary incomes. Additionally, they classify households according to the household heads' educational outcome, assuming that educational outcomes are correlated with lifetime income. 
Emphasizing challenges in finding a one-size-fits-all approach, \cite{shei2024distributional} propose that to use expenditure as an income proxy for groups that have low current earnings (but higher lifetime earnings) and current income for others. \cite{missbach2024distributional} remove households with total expenditures exceeding the mean total expenditure by five standard deviations. \cite{grainger2010pays} and \cite{linden2024manyfaces} excluding households with very high expenditure-to-income ratios. To avoid introducing bias by excluding households sharing a common characteristic, \cite{grainger2010pays} define income groups before dropping households.  

Concerned with fluctuations in expenditure patterns along the lifetime and resulting misrepresentations of lifetime incomes, \cite{hassett2009incidence} adjust a lifetime measure of consumption to correct for predictable changes along the life course. They construct sub-samples based on educational attainment and compute a typical path of consumption for each sub-sample and adjust households' consumption by the ratio of household consumption to average consumption of the appropriate age and educational group.


A best practice may be to include both income concepts, as done in \cite{grainger2010pays, wier2005co2, douenne2020vertical, yusuf2015distributional, rausch2011distributional, hassett2009incidence, fremstad2019impact, shei2024distributional}, and to include sensitive checks using the approaches described above. \\

\textit{Alternative outcome metrics}

Alternative outcome measures may be more appropriate depending on the research question. Table \ref{Tab:otherMeasures} provides an overview of alternative measures used in the literature, including behaviourally adjusted welfare measures (Compensating and Equivalent Variation), explicit measures of inequality (Gini index, Atkinson index), progressivity (Suits and Kakwani index), and poverty (energy poverty head count and FGT index), and measures of horizontal inequalities. 


\FloatBarrier
\begin{table}[H] 
\centering
\scriptsize
\begin{threeparttable}
     \caption{List of distributional impact measures other than the carbon tax as a share of income}
      \label{Tab:otherMeasures}
        \begin{tabularx}{\textwidth}{l|X} 

        \hline
        Measure & Papers \\ \hline
        \underline{Behaviourally adjusted welfare measures} & ~ \\ 
        Equivalent Variation & \cite{landis2019cost, reanos2022measuring, creedy2006carbon, tovar2023benefits, moz2021winners, semet2024coordinating} \\
        Compensating Variation & \cite{saelim2019carbon, vandyck2014distributional, tiezzi2005welfare, brannlund2004carbon, eisner2021distributional, renner2018household, li2020welfare, mardones2024contribution, vandyck2021climate} \\ 
        Equivalent Loss & \cite{labandeira1999combining} \\ 
        \underline{Poverty measures} & ~ \\ 
        Change in energy or fuel poverty & \cite{berry2019distributional, giraudet2021policies, reanos2021fuel, bourgeois2021lump, campagnolo2022distributional, fragkos2021equity, tobben2023unequal, buchs21} \\ 
        Change in poverty rate & \cite{brenner2007chinese, yusuf2015distributional, renner2018poverty, ravigne2022fair, saelim2019carbon, yusuf2015distributional, mardones2024contribution, chepeliev2021distributional} \\ 
        Change in the poverty gap & \cite{brenner2007chinese, malerba2021mitigating} \\ 
        Foster-Greer-Thorbecke index & \cite{renner2018household, malerba2021mitigating} \\ 
        Impact on poor relative to other & \cite{dorband2019poverty, christis2019detailed} \\ 
        \underline{Inequality indexes} & ~ \\ 
        Change in Gini index & \cite{wier2005co2, kerkhof2008taxation, Antosiewicz22, rub2024inequality, ravigne2022fair, wier2003environmental, dissou2014can, fragkos2021equity, linden2024manyfaces, tovar2023benefits, yusuf2015distributional, saelim2019carbon, mardones2020economic, symons1994carbon, eisner2021distributional, reanos2022distributional, li2020welfare, malerba2021mitigating, mardones2024contribution} \\ 
        D9/D1 ratio & \cite{Antosiewicz22, mardones2020economic, malerba2021mitigating, jaccard2021energy} \\ 
        Atkinson index & \cite{cornwell1996carbon, rub2024inequality, symons1994carbon, dissou2014can, mardones2024contribution} \\ 
        Reynolds-Smolensky index & \cite{cornwell1996carbon, jacobs2022distributional, ravigne2022fair, dissou2014can, bursens2026bridging, linden2024manyfaces} \\ 
        \underline{Progressivity indexes} & ~ \\ 
        Suits index & \cite{feindt2021understanding, wier2005co2, jiang2014distributional, linden2024manyfaces, shei2024distributional, eisner2021distributional, berry2019distributional, jacobs2022distributional, jiang2017analyzing, phungrassami2019fossil, semet2024coordinating} \\ 
        Kakwani index & \cite{cornwell1996carbon, jacobs2022distributional, ravigne2022fair, jiang2017analyzing, dissou2014can, bursens2026bridging} \\ 
        \underline{Net impacts} & ~ \\ 
        Share of reform winners/losers & \cite{douenne2020vertical, fremstad2019impact, chepeliev2021distributional, semet2024coordinating, jacobs2022distributional, vogt2019cash} \\ 
        West and Williams progressivity measure & \cite{west2004estimates, winter2023carbon} \\ 
        \underline{Carbon tax incidence} & ~ \\
        (Net) Carbon tax incidence & \cite{callan2009distributional, renner2018household} \\ 
        \underline{Price indexes} & ~ \\
        Changes in the CPI & \cite{reanos2022measuring, saelim2019carbon, mardones2020economic, chepeliev2021distributional} \\ 
        \underline{Horizontal inequality} & ~ \\ 
        Standard dev. within income group & \cite{fremstad2019impact} \\ 
        Measures of horizontal inequality & \cite{steckel2021distributional, missbach2024cash} \\ 
        \hline \hline
    \end{tabularx}
\end{threeparttable}
\end{table}

\FloatBarrier

Progressivity measures, particularly the Suits index, are popular to assess the distributional outcomes of carbon taxation. When carbon taxes and revenue recycling are assessed jointly, these measures can give misleading results because the carbon tax may be regressive (progressive) while the revenue recycling scheme is progressive (regressive). \cite{west2004estimates} and \cite{winter2023carbon} therefore propose an alternative measure. This index is closely related to the Suits index, with the index of two tax changes simply being the sum of the index for both tax changes separately.\\

\textit{Horizontal Inequalities}
\\
In recent years, horizontal inequalities (inequalities between groups based on location, gender, ethnicity, etc) have attracted interest. The most widely studied form of horizontal inequality concerns differences across urban and rural households, and differences across household types (age, race, and gender, presence of children, and educational achievement), while differences in asset ownership (home tenure, heating system, and vehicle ownership) remain relatively unexplored (see Figure \ref{FIG:horizontal} of the supplementary materials). 

Recent studies point towards substantial differences in burdens across dimensions other than income, including household composition, location, and other socioeconomic and demographic characteristics. \cite{cronin2019vertical} and \cite{tovar2023benefits} highlight that horizontal inequalities may be larger than vertical inequalities and that revenue recycling may amplify horizontal inequalities. In France, \cite{douenne2020vertical} highlights that carbon pricing may be regressive or progressive depending on whether vertical or horizontal inequalities are considered. In Sweden, \cite{brannlund2004carbon} show that households in sparsely populated areas bear larger net carbon tax burdens when revenues are returned as public transport subsidies, with no significant differences across regions if revenues are not recycled. Recently, studies pay more attention to the differences in carbon tax burdens within income groups \citep{rausch2011distributional, berry2019distributional, flues2015distributional, reanos2022measuring, tovar2023benefits, bourgeois2021lump, zhang2019effects, douenne2020vertical}. 

A strand of this literature uses regression techniques to estimate the relationship between socioeconomic characteristics and carbon tax incidences or burdens and identify disproportionally burdened groups \citep{tiezzi2005welfare, reanos2021fuel, saelim2019carbon, farrell2017factors, tobben2023unequal, levay2021association}. In a recent working paper, \cite{missbach2024distributional} uses supervised machine learning on a large sample covering 88 countries to identify characteristics most predictive of carbon-intensive consumption patterns. They then identify country clusters based on the predictors of high carbon intensity of consumption.


\section{Meta-analysis}\label{metaanalysis}

\subsection{Modelling choices in the literature}
The literature using microsimulation modelling to assess the distributional impacts of carbon taxation is growing. Most studies included in this literature review were written after 2015 (65\%) (Figure \ref{FIG:peryear} of the Supplementary materials). 

The majority focus on a single country (84\%), with over half studying European countries (52\%). France, Ireland, and the USA are the most commonly studied countries (Figure \ref{FIG:mapSingle} of the Supplementary materials). China and Thailand are the most studied Asian countries, and the USA, Mexico, and Canada are the most studied American countries. There is a notable lack of studies focusing on Africa, and only a few studies on South America. Cross-country comparative studies most frequently cover EU countries (Figure \ref{FIG:multicountry}  of the Supplementary materials). Studies in North America often compare the impact across states, regions, or provinces. 

\begin{figure}[ht]
	\centering
				\caption{Modelling choices as a percentage of total studies.}

		\copyrightbox[]{\includegraphics[width=0.7\textwidth]{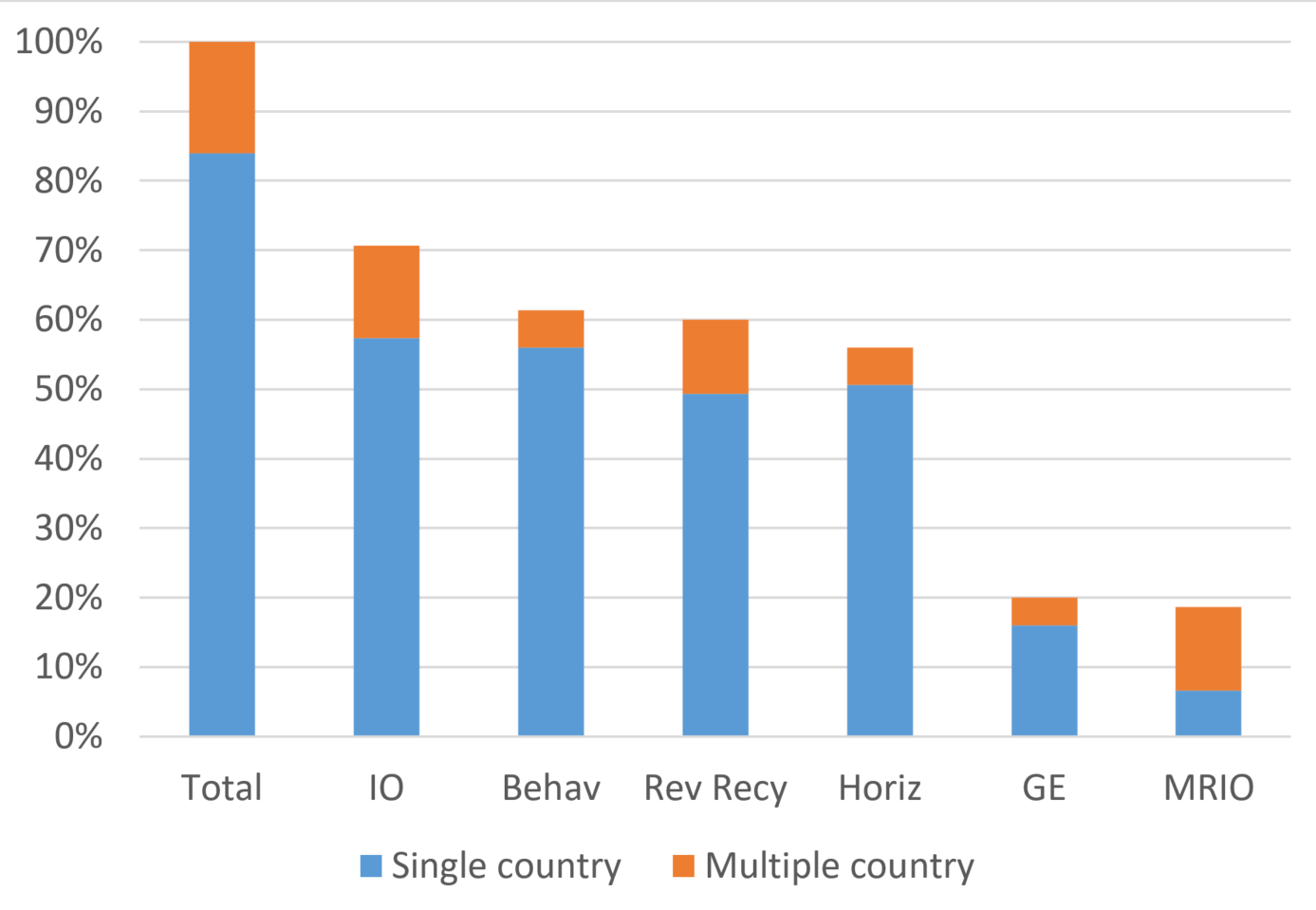}}%
		{}
		
	\label{FIG:method_choices}
\end{figure}

Figure \ref{FIG:method_choices} shows the percentage of studies using various modelling approaches. Most studies include indirect emissions through Input-Output models (70\%), household behavioural responses (62\%), some form of revenue recycling (60\%), or analysis of horizontal inequalities (55\%). Only a few recent studies utilize Multi-regional Input-Output models (20\%), most of which are cross-country comparative analyses. Few studies link microsimulation and GE models or include more than 10 heterogeneous household types (20\%).

\begin{figure}[ht]
	\centering
				\caption{Regressive and progressive impacts by developed and developing countries.}

		\copyrightbox[]{\includegraphics[width=0.7\textwidth]{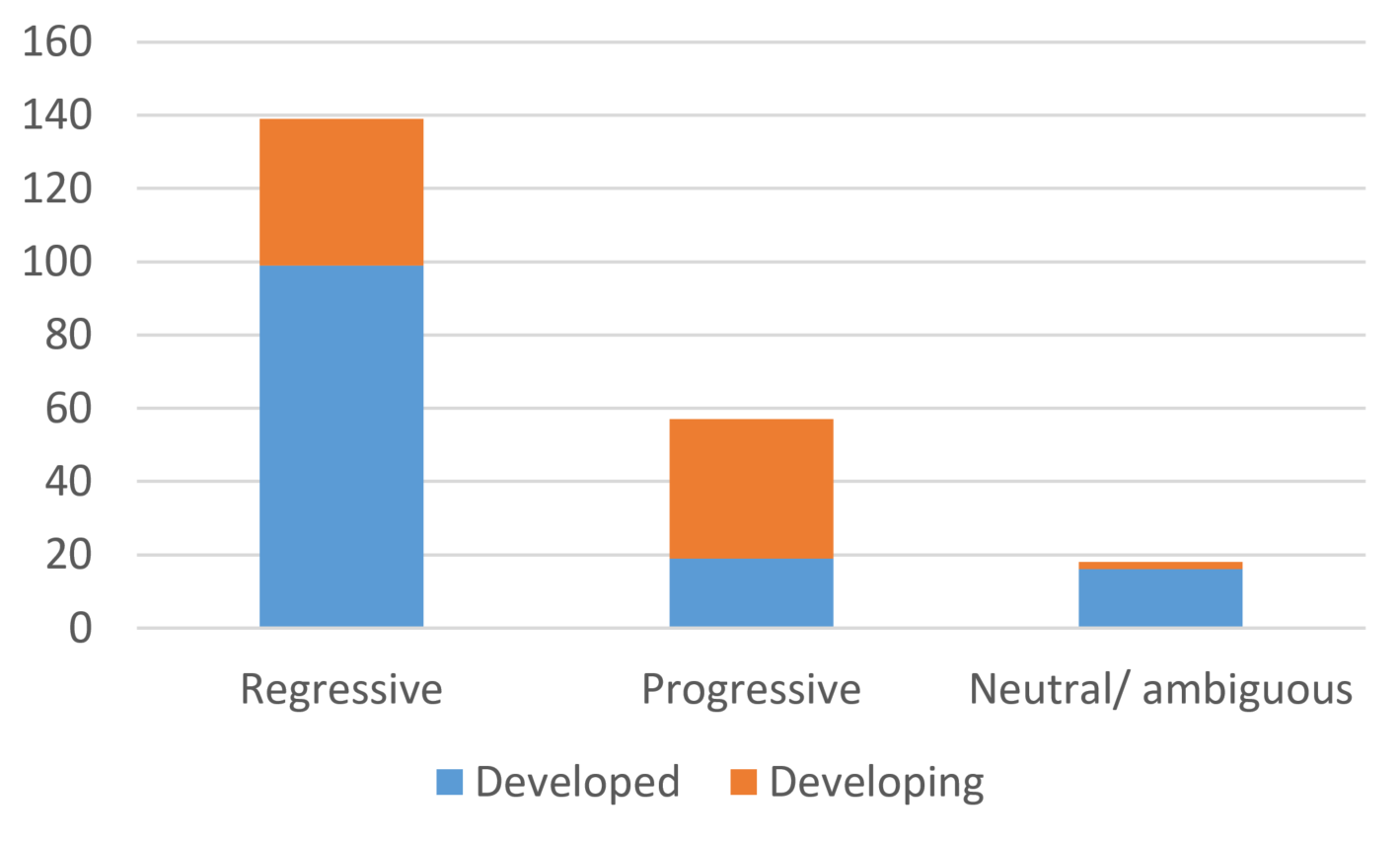}}%
		{}
		 \caption*{\footnotesize Country-level estimates. The total number of estimates included is 217. Where multiple estimates were available for one country, we included only the central scenario. The figure excludes estimates from \cite{dorband2019poverty}, \cite{jaccard2021energy}, and \cite{chepeliev2021distributional} because of difficulties in counting progressive or regressive results.} 
	\label{FIG:progressiveregressive}
\end{figure}

Figure \ref{FIG:progressiveregressive} counts regressive, progressive, and neutral or ambiguous estimates for developing and developed countries. In line with a previous review by \cite{ohlendorf2021distributional}, we find a larger share of progressive estimates in developing countries (44\% of estimates) than in developed countries (24\% of estimates).

\begin{figure}[ht]
	\centering
				\caption{Percent of Studies Finding Regressive Effects, by Modelling choice.}
 
		\copyrightbox[]{\includegraphics[width=0.7\textwidth]{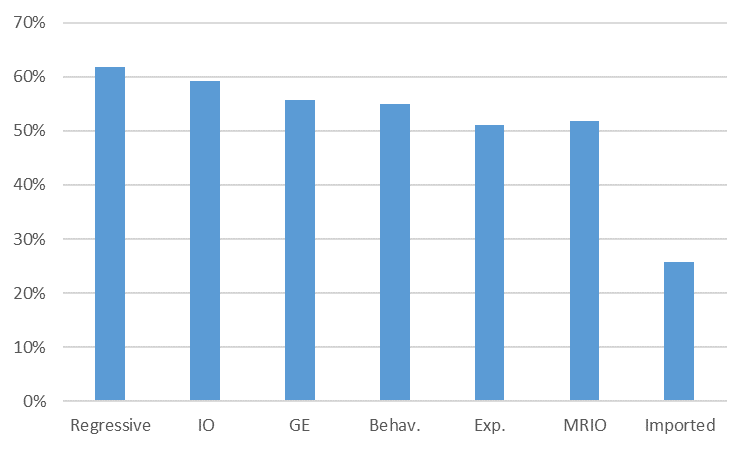}}%
		{}
		 \caption*{\footnotesize Values show regressive estimates as a percentage of estimates using the methodological approach. We show estimates before revenue recycling. Country-level estimates. The total number of estimates included is 217. Where multiple estimates were available for one country, we included only the central scenario. We exclude estimates from \cite{dorband2019poverty} and \cite{jaccard2021energy} because of difficulties in counting progressive or regressive results. GE = General equilibrium; IO = Input-Output; Exp. = Expenditure, MRIO = Multi-regional IO} 
	\label{FIG:modelling_prog}
\end{figure}

Figure \ref{FIG:modelling_prog} shows the percentage of regressive estimates by modelling approach. Approximately 62\% of included estimates are regressive. Among the estimates produced using GE models or those including household behavioural responses along the intensive margin, close to 55\% find regressive impacts, while the percentage lies close to 60\% for estimates produced with IO models. The share of regressive findings among studies using MRIO models or that use expenditure as the welfare concept lies just above 50\%. The lowest share of regressive results is reported with studies that simulate a tax covering imported indirect emissions at 26\%, compared to close to 60\% for all studies computing indirect emissions (i.e. those using IO models). 

However, we can not infer the impact of these modelling approaches on regressivity estimates from Figure \ref{FIG:modelling_prog}. Firstly, we pool estimates across country contexts, including developed and developing countries. Progressive estimates are more likely in developed countries \citep{ohlendorf2021distributional}, but developed countries are not equally represented across categories\footnote{30\% of estimates are from developing countries, whereas 10\% of studies using GE are set in developing countries, 36\% studies using IO and 34\% of studies using expenditure as welfare concept are set in developing countries.}. Differences across categories could result from over- or under-representation of developing countries. Secondly, the figure only includes GE studies that explicitly link GE to microsimulation models or that introduce more than 10 heterogeneous agents, excluding a significant share of GE studies. To address some of these issues, the next section provides regression-based estimates of the impact of modelling choices on distributional impact estimates.

\clearpage

\subsection{Probit model}

What are the implications of the modelling choices and their implementation on distributional impact estimates? Using a probit model, this section estimates the isolated influence of these choices on distributional impact estimates. We go beyond the broad conceptual choices investigated by \cite{ohlendorf2021distributional} and test whether the implementation of some key modelling choices affects estimates. We base our approach on \cite{greene2012201}. 


We use a probit model to estimate the relationship between a binary dependent variable and a set of independent variables. In this case, the binary outcome represents the likelihood of regressivity (0 = progressive; 1 = regressive). We classify study results as regressive or progressive using authors' statements about the estimated distributional impact (before revenue recycling). To ensure a sufficient sample size for meaningful analysis, we aggregate studies reporting neutral distributional effects with those indicating progressive impacts. Estimated coefficients reflect how changes in the independent variables are associated with changes in the probability of observing a regressive outcome. 

Let \( y_i \) be a binary indicator equal to 1 if study \( i \) finds that carbon pricing is regressive, and 0 otherwise. The probit model assumes the existence of a latent variable \( y_i^* \) such that:
\[
y_i^* = X_i \beta + \varepsilon_i, \quad \varepsilon_i \sim \mathcal{N}(0,1)
\]
\[
y_i = 
\begin{cases}
1 & \text{if } y_i^* > 0 \\
0 & \text{otherwise}
\end{cases}
\]

The probability that a regressive outcome is reported is given by:
\[
\Pr(y_i = 1 \mid X_i) = \Phi(X_i \beta)
\]
where \( \Phi(\cdot) \) is the standard normal cumulative distribution function. Coefficients \( \beta \) are estimated via maximum likelihood. For interpretability, we also report average marginal effects, which represent the change in the probability of a regressive outcome associated with a one-unit change in each covariate.


We include 217 estimates from 71 countries\footnote{Including regions from \cite{chepeliev2021distributional}; Rest of EU and EFTA, Rest of Europe, Rest of Central Asia, Rest of East Asia, Rest of South Asia, North Africa, Sub-Saharan Africa.} and for 30 different years. We test the impact of conceptual choices on the distributional impact estimates, including general equilibrium effect, indirect effects, household behavioural responses, and the use of expenditure to approximate lifetime income, similar to \cite{ohlendorf2021distributional}. We go beyond \cite{ohlendorf2021distributional} by investigating the impact of the implementation of such choices, including the use of a MRIO, the use of GTAP as MRIO, the coverage of imported emissions, the welfare or inequality metric use, the data year, and coverage of GHG emissions other than $CO_2$. Control variables include GDP per capita, the rural population share, the Gini index, and the poverty gap at \$6.85 per day (2017 PPP) and are sourced from the World Bank matching the data year to the studies' microdata (HBS) year. 

We also conduct multiple sensitivity checks. We impose cluster-robust standard errors by country to address the non-independence of observations. We also cluster standard errors by study to test the sensitivity of our estimates to the clustering decision. We include region-income fixed effects to address concerns that results from one country or region may drive our estimates. We also test the sensitivity of our results to the inclusion of studies using GE models, because we restrict the sample to studies with an explicit micro-macro linkage and may therefore introduce selection bias. Lastly, we use jackknifing to test the impact of single countries on our results.  

\subsubsection{Results}


Table \ref{tab:probit} shows the results of our regression analysis for multiple models. In column (1), we begin with a model that tests the impact of conceptual modelling choices on distributional impact estimates, approximately replicating \cite{ohlendorf2021distributional}'s specification. We find statistically significant associations between the inclusion of demand-sided responses, the age of the data, and the use of expenditure as a proxy for lifetime income with less regressive estimates. Similarly, including indirect effects through IO models, GE effects, and a developing country context increases the likelihood of less regressive findings, though these impacts are statistically insignificant. These findings are in line with \cite{ohlendorf2021distributional}.

Adding variables capturing the implementation of conceptual modelling choices increases the models' explanatory power from a Pseudo $R^2$ of 0.254 (column 1) to 0.394 (column 4). Further, including variables capturing the implementation of conceptual modelling choices renders the association between lifetime income and progressive estimates insignificant, and changes the sign of the GE and indirect effects variables. Column (4) shows that the use of IO models is insignificantly associated with more regressive estimates, but that the inclusion of MRIO, the use of the GTAP database, exempt fuels or sectors, and the coverage of imported emissions is associated with more progressive outcomes, while coverage of other GHG emissions is associated with more regressive impacts. Among these variables, only the inclusion of imported emissions is statistically significant and strongly impacts estimates. 

Once we account for alternative welfare concepts and differences in tax coverage, the significant impact of lifetime income also disappears, though a negative association between lifetime income and regressivity persists. The negative association between the inclusion of demand-sided responses and regressivity is significant, independent of the model used. Finally, we find that using inequality or progressivity indicators other than the variation of the carbon tax as a share of income is associated with a higher likelihood of finding progressive impacts. 

To provide a more intuitive interpretation of the results, Table \ref{tab:ordered_probit_margins} shows the marginal effects of Table \ref{tab:probit}. Focusing on model (4), the largest significant impacts on the probability of finding regressive results are found for the inclusion of imported emissions, demand-sided responses, and the use of older data and explicit inequality or regressivity metrics. Including imported emissions, demand-sided responses, using older HBS data, and using explicit inequality and progressivity measures are associated with a lower probability of finding regressive results, conditional on the other modelling choices in model (4) (39.4 percentage points, 22.4 percentage points, 1.2 percentage point per year, 17.7 percentage points). The only variable that is significantly associated to a higher likelihood of finding regressive results is the inclusion GE effects, which increase this likelihood by 22.3 percentage points condiitonal on all other modelling choices.  

One caveat should be kept in mind when interpreting these results. First, our model only accounts for some differences across models. Many differences are unaccounted for, including differences in tax coverage (beyond the emission coverage and the presence of some exemptions), the expenditure system used to model household behaviour, the level of heterogeneity in response across households, the unit of analysis (household or individual), the ranking variable used, the granularity of household energy expenditure, and the specific measure of inequality and progressivity. These nuanced differences likely affect distributional impact estimates, but are not considered in our probit model. 

\clearpage
\small	
\begin{ThreePartTable}
\begin{longtable}{p{0.25\textwidth}ccccc}
\caption{Probit results - Impact of modelling choices on distributional impact estimates} \label{tab:probit} \\ 
\toprule
\textbf{Regressive = 1} & \textbf{(1)} & \textbf{(2)} & \textbf{(3)} & \textbf{(4)} \\ 
\midrule
\endfirsthead
\caption[]{(continued)} \\ 
\toprule
\textbf{Distr. Impact} & \textbf{(1)} & \textbf{(2)} & \textbf{(3)} & \textbf{(4)} \\ 
\midrule
\endhead
\midrule
\multicolumn{5}{r}{\textit{Continued on next page}} \\ 
\midrule
\endfoot
\bottomrule
\endlastfoot
\underline{Data Year}  &  &  &  &  \\

HBS Data year & - 0.013         &      -0.049\sym{**} &      -0.038\sym{*}  &      -0.056\sym{**} \\
            &     (0.022)         &     (0.024)         &     (0.023)         &     (0.024)         \\ \\

\underline{Tax coverage}  &  &  &  &  \\
IO & -0.062         &       0.364         &      -0.014         &       0.172         \\
            &     (0.501)         &     (0.490)         &     (0.471)         &     (0.512)         \\
MRIO  &                     &       0.468         &       0.811         &       0.595         \\
            &                     &     (0.718)         &     (0.692)         &     (0.728)         \\
GTAP &                     &      -1.070         &      -1.149         &      -1.024         \\
            &                     &     (0.960)         &     (0.831)         &     (0.988)         \\
Imported emissions &                     &      -1.700\sym{**} &      -1.524\sym{***} &      -1.842\sym{***}\\
            &                     &     (0.660)         &     (0.505)         &     (0.696)         \\
Exemptions &                     &      -0.439         &       0.069         &      -0.352         \\
            &                     &     (0.593)         &     (0.524)         &     (0.540)         \\
Other GHG  &                     &       0.186         &       0.061         &       0.232         \\
            &                     &     (0.509)         &     (0.503)         &     (0.517)         \\
 
\underline{Behaviour}  &  &  &  &  \\
Demand-sided &      -1.009\sym{**} &      -1.035\sym{***}&      -0.766\sym{**} &      -1.047\sym{**} \\
            &     (0.449)         &     (0.385)         &     (0.360)         &     (0.407)         \\
GE &      -0.189         &       1.017\sym{**} &       0.788\sym{*}  &       1.042\sym{**} \\
            &     (0.563)         &     (0.461)         &     (0.461)         &     (0.472)         \\
 
\underline{Welfare concept}  &  &  &  &  \\
Life-time income &      -0.795         &      -0.018         &      -0.061         &       0.075         \\
            &     (0.575)         &     (0.424)         &     (0.376)         &     (0.423)         \\
Behav. adj. welfare  &                     &       0.255         &       0.469         &       0.381         \\
            &                     &     (0.410)         &     (0.435)         &     (0.423)         \\
Ineq. or progr. measure &                     &      -0.820\sym{**} &      -0.685\sym{*}  &      -0.822\sym{**} \\
            &                     &     (0.377)         &     (0.350)         &     (0.344)         \\

\underline{Contextual variable}  &  &  &  &  \\
Developing  & -0.016         &                     &      -0.169         &      -0.270         \\
            &     (0.325)         &                     &     (0.492)         &     (0.550)         \\

 &  &  &  &  \\ 

Other control variables & Yes & No & Yes & Yes \\
Region-income-fixed effects & Yes & Yes & No & Yes \\
Paper-clustered st. error & Yes & Yes & Yes & Yes \\ \hline


 Observations  &         199         &         212         &         199         &         199         \\
 Pseudo $R^2$ &   0.202 &  0.413 &  0.371 &  0.408 \\ \hline

\end{longtable}
\end{ThreePartTable}
\footnotesize Note: Distr. Impact: 0 = progressive, 1 = regressive. Higher values of HBS Data year indicate older datasets. We include 9 region-income groups, differentiating between 3 geographical regions and 3 income levels. Low-and middle-income countries in Europe and middle-and high-income countries in Asia, Africa, and Australia were grouped into two separate groups to increase the observations in the fixed-effect category. Other control variables include GDP per capita, Gini index, poverty gap at \$6.85/day (2017 PPP), rural population share, and dummy variables for low and middle-income countries. Standard errors in parentheses. \\ 
\footnotesize *** p$<$0.01, ** p$<$0.05, * p$<$0.1

\clearpage
\small	
 
\clearpage

\begin{ThreePartTable}
\begin{longtable}{p{0.25\textwidth}ccccc}
\caption{Impact of modelling choices on distributional impact estimates - Average marginal effects} \label{tab:ordered_probit_margins} \\ 
\toprule
\textbf{Regressive = 1} & \textbf{(1)} & \textbf{(2)} & \textbf{(3)} & \textbf{(4)} \\ 
\midrule
\endfirsthead
\caption[]{(continued)} \\ 
\toprule
\textbf{Regressive = 1} & \textbf{(1)} & \textbf{(2)} & \textbf{(3)} & \textbf{(4)} \\ 
\midrule
\endhead
\midrule
\multicolumn{5}{r}{\textit{Continued on next page}} \\ 
\midrule
\endfoot
\bottomrule
\endlastfoot
\underline{Data Year}  &  &  &  &  \\

HBS Data year &     -0.004         &      -0.011\sym{**} &      -0.009\sym{*}  &      -0.012\sym{**} \\
            &     (0.007)         &     (0.005)         &     (0.005)         &     (0.005)         \\

\underline{Tax coverage}  &  &  &  &  \\
IO &   -0.018         &       0.079         &      -0.003         &       0.037         \\
            &     (0.149)         &     (0.105)         &     (0.107)         &     (0.109)         \\
MRIO  &                  &       0.101         &       0.184         &       0.127         \\
            &                     &     (0.156)         &     (0.155)         &     (0.156)         \\
GTAP &                      &      -0.232         &      -0.260         &      -0.219         \\
            &                     &     (0.207)         &     (0.185)         &     (0.210)         \\
Imported emissions &                 &      -0.368\sym{***}&      -0.345\sym{***}&      -0.394\sym{***}\\
            &                     &     (0.136)         &     (0.106)         &     (0.142)         \\
Exemptions &                &      -0.095         &       0.016         &      -0.075         \\
            &                     &     (0.128)         &     (0.119)         &     (0.115)         \\
Other GHG  &                 &       0.040         &       0.014         &       0.050         \\
            &                     &     (0.110)         &     (0.114)         &     (0.111)         \\
 
\underline{Behaviour}  &  &  &  &  \\
Demand-sided &      -0.299\sym{**} &      -0.224\sym{***}&      -0.173\sym{**} &      -0.224\sym{***}\\
            &     (0.129)         &     (0.079)         &     (0.081)         &     (0.084)         \\
GE &      -0.056         &       0.220\sym{**} &       0.178\sym{*}  &       0.223\sym{**} \\
            &     (0.166)         &     (0.097)         &     (0.102)         &     (0.098)         \\
  
\underline{Welfare concept}  &  &  &  &  \\
Life-time income &      -0.236         &      -0.004         &      -0.014         &       0.016         \\
            &     (0.156)         &     (0.092)         &     (0.085)         &     (0.091)         \\
Behav. adj. welfare  &                  &       0.055         &       0.106         &       0.082         \\
            &                     &     (0.088)         &     (0.098)         &     (0.090)         \\
Ineq. or progr. measure &                 &      -0.178\sym{**} &      -0.155\sym{**} &      -0.176\sym{**} \\
            &                     &     (0.076)         &     (0.077)         &     (0.070)         \\

\underline{Contextual variable}  &  &  &  &  \\
Developing  &   -0.005         &                     &      -0.038         &      -0.058         \\
            &     (0.096)         &                     &     (0.111)         &     (0.119)         \\

 &  &  &  &  \\ 

Other control variables & Yes & No & Yes & Yes \\
Region-income-fixed effects & Yes & Yes & No & Yes \\
Paper-clustered st. error & Yes & Yes & Yes & Yes \\ \hline


 Observations  &         199         &         212         &         199         &         199         \\
 Pseudo $R^2$ &   0.202 &  0.413 &  0.371 &  0.408 \\ \hline

\end{longtable}
\end{ThreePartTable}
\footnotesize Note: Distr. Impact: 0 = progressive, 1 = regressive. Higher values of HBS Data year indicate older datasets. We include 9 region-income groups, differentiating between 3 geographical regions and 3 income levels. Low-and middle-income countries in Europe and middle-and high-income countries in Asia, Africa, and Australia were grouped into two separate groups to increase the observations in the fixed-effect category. Other control variables include GDP per capita, Gini index, poverty gap at \$6.85/day (2017 PPP), rural population share, and dummy variables for low and middle-income countries. Standard errors in parentheses. \\ 
\footnotesize *** p$<$0.01, ** p$<$0.05, * p$<$0.1

\clearpage

\normalsize

\subsection{Modelling challenges}

Our meta-analysis demonstrated that modelling choices and differences in their implementation significantly affect the qualitative conclusions of distributional impact assessments, particularly the inclusion of household behavioural responses, coverage of imported indirect emissions, and the outcome metric used to assess distributional impacts. While we attempted to include important drivers of the distributional impact in previous sections, various relevant drivers and modelling choices have not been discussed yet. This section briefly reviews these modelling choices and discusses challenges for modellers in incorporating them. \\

\textit{Cost Pass-through to consumers}

When a (carbon) tax is imposed on products, the tax may only be partially passed through to consumers, with the remainder absorbed by firms. Conversely, the pass-through may sometimes be larger than 1, as firms increase prices beyond the cost of the tax as a result of imperfectly competitive product markets \citep{ganapati2020energy}.

Environmental microsimulation models commonly assume full pass-through of the carbon tax from producers to consumers. Cost pass-through or forward cost-shifting may however be a major determinant of carbon pricings' distributional impact \citep{fullerton2008distributional, shang2023poverty}, particularly if it is uneven across products, but it has been largely ignored in microsimulation modelling. 

While the empirical evidence of carbon tax pass-throughs is still emerging, the assumption of full pass-through rates may be reasonable in the short term. \cite{fabra2014pass} find that the EU-ETS price is fully passed through to consumers under the form of higher electricity prices. Similarly, \cite{bernard2021impact} find full pass-through of the Canadian carbon tax. \cite{ganapati2020energy} find an average pass-through rate of 70\%, though with substantial heterogeneity across sectors, ranging from 0.36 for gasoline to 1.87 for cement. \cite{bovenberg2001neutralizing} show that a carbon tax increases the price of coal by 90\% of the total tax. A similar result is found in \cite{metcalf2008analysis}.

Faced with substantial uncertainty around pass-through rates, \cite{akouguz2020new} propose that the pass-through rate in a partial equilibrium framework could be assumed to be given by the ratio of market demand and supply elasticities, $\theta=\eta^S/(\eta^S-\eta^D)$, where $\eta^S$ and $\eta^D$ are the market demand and supply elasticities and $\theta$ is the pass-through rate. Therefore, the more (in)elastic demand (supply), the lower (higher) the pass-through rate. Alternatively, \cite{hamilton1994simulating} use the share of exports in the total production of each sector to scale the pass-through rate. \\

\textit{Unit of analysis}

The unit of analysis of most distributional impact assessments is the household. This may impact estimates of the distributional impact, similarly to how using equivalent scales can change distributional impact estimates by altering the position of households along the income distribution \citep{cowell1999equivalence}. For example, \cite{grainger2010pays} show that using equivalence scales or the individual (rather than the household) as a unit of analysis can lead to more regressive estimates. \\

\textit{Energy prices and quantity estimates}

Carbon taxes are levied relative to the quantity of $CO_2$ emissions associated with the consumption of the good or service. Expenditure and Input-Output data, however, commonly report consumption in monetary units. It is therefore often necessary to transform expenditures into quantities using price information. Prices of energy goods in particular vary significantly over time and by quantity purchased, contract, and source (import or domestically produced). This implies that a given monetary amount may reflect different quantities of energy commodities purchased. It may, therefore, aid transparency to report energy price data alongside the carbon intensity factors. \\

\textit{Translation between industrial and consumer good classifications}

Combining IO models with microsimulation models requires translating across statistical classifications (from COICOP product classification to ISIC or NACE industry classifications) using bridging matrices. Some MRIO databases, notably GTAP, provide such tables directly. For EU countries, tables are provided by \cite{cai2020bridging}. \cite{cazcarro2022linking} develop the approach taken by \cite{cai2020bridging} and provide a guide to constructing bridging matrices. Translating between both classifications requires assumptions about the structure of industry outputs. A common assumption is the Fixed Product Sales Structure Assumption, which assumes that all products produced within an industry are produced using the same input. In some sectors, such as agriculture, this assumption can be problematic, and some analyses may require alternative approaches to differentiate between products within the industry (e.g. food carbon footprints). \\

\textit{Aggregation levels and product heterogeneity}

Issues may arise because products with widely different environmental impacts are grouped within the same expenditure or industry category (e.g. the Petroleum industry produces fuel, plastics and chemical feedstock). MRIO databases often differ in their level of sector aggregation. Relative to WIOD, which includes 56 sectors and 49 countries (including a category for the rest of the world), GTAP 11 (65 sectors), EXIOBASE 3.6 (163 sectors), Eora (up to 15,909 sectors) has a higher sectorial resolution but relies heavily on imputation methods \citep{lenzen2011aggregation}. Further differences are discussed in \cite{odonoghue2024distributional}. This issue can be addressed by separating the groups using benchmark IO files or aggregate statistics sourced elsewhere (for example \cite{hassett2009incidence}). 

A comparable issue can arise when expenditure survey data is provided at a high level of aggregation (grouping bread and beef, or gas, electricity, and fuel oil). \cite{levay2021association} addresses this issue by estimating a regression model of energy consumption elsewhere and splitting the domestic energy category using estimated coefficients. 

An additional challenge arises from the absence of price or quantity information in most expenditure surveys. As household budget and expenditure surveys report expenditures but not quantities consumed, it is impossible to differentiate between budget and luxury goods. As carbon taxes are levied in relation to quantities and researchers commonly apply one emission coefficient to households, this likely leads to overestimated (underestimated) carbon tax burdens for high (low) income households. This issue is discussed at length in \cite{andre2024challenges}.\\

\textit{Linear price elasticities with respect to price}

A challenge in modelling household behaviour is that demand elasticities are estimated linearly with respect to prices, i.e. the estimated elasticity does not vary with the initial price level and the size of the price change. Responses may be small or non-existent if the price increase is small, and much larger when the price change is large. There may therefore be a "threshold of pain" beyond which consumers react more strongly. Non-linear responses with respect to price are acknowledged in \cite{mardones2024contribution}, which uses average and high elasticities under a carbon tax of 5, 25, and 50 USD. Overall, the literature, however, largely abstracts from this issue. \\

\textit{Modelling the extensive margin and complementary goods}

Modelling household behaviour along the extensive margin, i.e. whether a household consumes a certain commodity or not, is challenging for two reasons. Firstly, expenditure survey data typically do not include contextual information relevant to technology adoption and other changes along the extensive margin. Relevant information includes information on household assets and their characteristics, dwelling characteristics, location, public transportation access, and home ownership. Second, household adoption behaviour is driven by a multitude of factors, including economic considerations, needs, social norms, attitudes, product characteristics, and the diffusion level of the technology \citep{ajzen1991theory, rogers2010diffusion, zilberman2011consumption, david1969contribution}. Only a few studies model the extensive margin \citep{bourgeois2021lump, giraudet2021policies, ravigne2022fair}. These studies, however, do not model the adoption decision but simulate the distributional impact of technology adoption under assumed diffusion patterns in line with national targets. \\

\textit{Timeliness of the data}

Household expenditure data and the IO models used in constructing models are frequently a few years old. Consequently, households may have adjusted their consumption patterns, and firms may have changed their input structure and improved their energy efficiency. Further, the demographic structure, income levels, and prices may have changed substantially since the data was collected. One avenue that remains unexplored in the literature is to use nowcasting and calibration techniques to recalibrate output, employment, and energy-and carbon intensity at the industry level. This approach proved useful during the COVID-19 crisis \citep{sologon2022covid, barbaglia2023testing, o2020modelling}. Developing and validating nowcasting models is costly and involve a series of data adjustments that can be problematic when the data is used as input for further modelling. 

Therefore, \cite{IMMERVOLL2025114783} take a simpler approach, arguing that extensive data manipulations can obscure distributional results from the policy modelling that are of primary interest. To account for demographic, price, and income changes between the data year and the introduction of the analysed carbon tax in Lithuania, \cite{IMMERVOLL2025114783} uprate household incomes and product prices using decile-level income inflators and Harmonized Index of Consumer Prices (HICP), and apply static reweighting to reflect changes in the population between 2015 and 2022. This simple approach addresses some issues related to outdated data, but ignores important changes such as changes in households' consumption behaviour and firms' energy mix and supply chains.\\

\textit{Expenditure survey limitations}

Household expenditure surveys suffer from various known limitations; they do not include quantity and asset ownership information, do not capture in-kind service consumption or own-production, insufficiently cover the tails of the income distribution, and suffer from under-reporting of certain purchases (such as tobacco, fuel, and alcohol) and infrequent purchases. 

Under-reporting and missing information can be addressed through imputation. \cite{brannlund2004carbon} address under-reporting of infrequent purchases by following \cite{heckman1979sample}, and estimating a Tobit model and a truncated regression for each expenditure group, following a two-step procedure. \cite{levay2021association} smooth out infrequent expenses over household clusters, assigning equal shares of the cluster-level expense to all households within the cluster using a Heckman selection model to address missing fuel use and ownership of company cars. \cite{missbach2024distributional} address misreporting by excluding households if their expenditure on an item is in the 99th percentile of all non-zero expenditures and replacing it with the median level.\\

\textit{Sensitivity analysis}

Conceptual modelling choices and their implementation affect distributional impact assessments. Modellers are therefore well-advised to test the sensitivity of their results to changes in the conceptual modelling components, and where possible, their implementation. Some authors include behavioural responses as a sensitivity analysis \citep{feindt2021understanding, mathur2014distributional, linden2024manyfaces}. \cite{feindt2021understanding} borrow estimates of own-price elasticities from the literature,  while \cite{linden2024manyfaces} estimate a linear expenditure system. \cite{fremstad2019impact} vary the strength of the behavioural response along the income distribution, showing three scenarios. Others vary the tax coverage. \cite{feindt2021understanding} simulate a scenario where all domestic EU emissions are covered (rather than all emissions worldwide). \cite{linden2024manyfaces} include a scenario where all emissions worldwide are covered (rather than domestic emissions only). \cite{fremstad2019impact} test the sensitivity of their results to their carbon intensity estimates by using an alternative carbon intensity factors. Other studies include different rates of pass-through \citep{mathur2014distributional}, adjustments for underreporting \citep{feindt2021understanding}, and rebound effects \citep{jacobs2022distributional} as sensitivity checks. \\

\section{Directions for future research}\label{FurtherResearch}
While the literature on distributional impact assessments of carbon taxation is growing, some areas remain relatively unexplored. Key topics include the simulation of actual policy, behavioural responses to revenue recycling, household technology adoption, interactions with tax-benefit systems, impacts on the wealth distribution, and co-benefits and the cost of inaction.

A first area for future research concerns modelling actual carbon pricing schemes. In practice, carbon prices differ substantially across sectors and fuels, with differences in rates and coverages across countries \citep{OECD2021ECR}. Exemptions and differentiated carbon prices alter the distributional impact of carbon pricing \citep{renner2018household}. Recently, the OECD and the World Bank developed methodologies to account for differences in coverage and rates \citep{OECD2021ECR, agnolucci2024measuring}. A chapter in the 2024 OECD employment outlook studies the change in the distributional impact of effective carbon rates (ECRs) over the years for a selection of countries \citep{elgouacem2024pays}. Future research could expand on this work. 

Future research could also consider interactions with tax-benefit and welfare systems, such as VAT knock-on effects, the role of revenue recycling in addressing social protection gaps, or impacts on prices and the real value of benefits. While microsimulation models are apt to model tax-benefit systems, the interactions between tax-benefit systems and carbon taxes remain unexplored. 

Further research could also focus on the impact of revenue recycling packages. Few studies consider policy packages and combinations of policies \citep{symons1994carbon, Antosiewicz22, ravigne2022fair, van2024public}. Revenue recycling packages may address equity concerns whilst maintaining firm competitiveness \citep{vandyck2021climate}, or manage the equity-effectiveness trade-off\footnote{Redistributing to the poor reduces the effectiveness of the carbon price because the poor have the most carbon-intensive consumption baskets and higher propensities to consume fuel.} of carbon taxation and revenue recycling \citep{sager2019income, tovar2023benefits, semet2024coordinating}, including rebound effects \citep{ravigne2022fair, bourgeois2021lump, giraudet2021policies}. 

As discussed above, modelling household behaviour along the extensive margin is challenging and rarely considered in distributional impact assessments. Only a few studies integrate changes along the extensive margin in their assessments. These studies typically rely on specialized models and projections (RES-IRF model used by \cite{bourgeois2021lump} and \cite{giraudet2021policies} or IMACLIM-3ME and MATISSE\footnote{the MATISSE model - Microsimulation Assessment within the low-carbon Transition of Inequalities and Sustainable Systems of Energy - is open source and its R code can be downloaded from https://github.com/eravigne/matisse} used by \cite{ravigne2022fair}). Future research could focus on modelling behaviour along the extensive margin for distributional impact assessment. For a review of theoretical and econometric models of discrete-continuous choices with application to residential energy demand, we refer to \cite{hanemann2024discrete}. 

A challenge in measuring the distribution impact of green subsidies and behaviour change along the extensive margin, and for the assessment of distributional impacts of carbon taxation more broadly, is the absence of asset and wealth data in most expenditure datasets used in the literature. Without information on real estate ownership and indirect incomes, car ownership, savings stocks, and access to credit markets, it is difficult to micro-found models of innovation adoption and energy efficiency investments. Future research could focus on integrating wealth and expenditure data to address this limitation. We are only aware of one study that combines wealth and expenditure data to study carbon footprints along the wealth distribution \citep{buchs2024emission}.

Relatedly, we are not aware of any microsimulation studies that investigate the impact of a carbon tax and revenue recycling on wealth. 

Lastly, co-benefits on health, biodiversity, asset value, increased comfort, and well-being, the monetary value of avoided health and economic damages (i.e. the benefits of climate change mitigation) are rarely included. These effects are, however, central dimensions of the distributional impact of all climate change mitigation policies \citep{drupp2025economics}, and can impact estimates substantially, as shown by \cite{tovar2023benefits} who find reductions in regressivity when avoided economic damages are included.

\section{Conclusion}
Distributional impacts of carbon taxation receive substantial attention in policy discussions, academia, and the media. These impacts are frequently assessed ex-ante using bottom-up (micro)simulation models. The choices made in developing these models affect the distributional impact estimates and can lead to divergent results. Previous research focused on the impact of various study designs (here referred to as conceptual choices) on distributional impact estimates.

This paper makes two contributions to the literature. First, it provides an in-depth review of the components of environmental microsimulation models of carbon taxation and takes stock of the approaches in which model components have been implemented. Second, it conducts a meta-analysis to estimate the impact of the inclusion of model components (conceptual choices) and their implementation (implementation choices) on the likelihood of finding regressive distributional impacts of carbon taxation. 

Using a sample of 217 estimates across 71 countries, our meta-analysis demonstrates that the modelling choices and the implementation of such choices matter. The explanatory power of our probit model increases by 55\% once we include predictors of differences in the implementation of conceptual choices, not considered in previous work. Despite differences in the scope and sample, our results confirm that the inclusion of household behavioural responses reduces carbon tax regressivity, but also show that the link between carbon tax regressivity and indirect emission coverages found by \cite{ohlendorf2021distributional} is mediated by the coverage of imported indirect emissions. This has important implications for carbon tax design, as it suggests that taxes covering indirect (embedded) imported emissions, for example through a Carbon Border Adjustment Mechanism, would render the distributional impact of carbon taxation less regressive. Further, our analysis shows that, beyond coverage of imported indirect emissions, the use of older expenditure data sets, modelling demand-sided responses, and using explicit measures of inequality and progressivity make it significantly more likely to find progressive impacts. Lifetime income proxies, multi-regional IO models, the use of the GTAP database, and sector exemptions are also associated with more progressive outcomes, though these effects are statistically insignificant. Similarly, coverage of GHG emissions other than $CO_2$, using behaviourally adjusted welfare metrics and using MRIO models is insignificantly associated with more regressive estimates. Lastly, we  find that the only variables that is significantly associated with more regressive estimates is the inclusion of GE effects.

Another central contribution of this paper relates to an in-depth survey of the modelling landscape. Despite the growing popularity of these models, no standard practices are emerging. We provide the most in-depth discussion of conceptual choices and the first stock-take of the approaches to implementing conceptual choices in microsimulation models. With the stock-take of modelling approaches and their implementation, we hope to contribute towards the diffusion of standard practices, aid the transparency and interpretability of model outputs, and facilitate training modellers and analysts. 

Establishing standard practices would aid the comparison of impacts across countries and carbon tax designs, and facilitate drawing general insights from the literature. Nonetheless, different conceptual and methodological approaches can be appropriate depending on the context, available data, and the research question. There would, therefore, likely be benefits to establishing a knowledge base and platforms for cooperation. Establishing a knowledge base as a public good for the research and policy community through the collection of standardized modelling approaches, open-source models, data banks of parameters\footnote{One example of a collection of carbon tax incidence estimates is provided here \url{https://www.cpic-global.net/}} and algorithms would aid cooperation and the production of comparable results across multiple research teams, support transparency, and facilitate learning. 

The last sections of this paper highlighted key challenges for modellers and identified areas for future research. We show that the EU and the U.S. are studied most, while other geographical areas, notably African countries, are rarely studied. Where these countries are included, they are analysed in large samples of countries and detailed analysis is often missing \citep{dorband2019poverty, missbach2024distributional}. Distributional impacts of clean technology adoption, actual (rather than hypothetical) policies, costs of inaction, and co-benefits of action remain largely unexplored.


\clearpage
\appendix

\section*{Appendix A: Supplementary Material}

\subsection{Study sample}

\begin{longtable}{ p{.45\textwidth}  p{.45\textwidth} } 
   \caption{List of papers retained from the literature search} \label{listofstudies}
    \endfirsthead
    \hline
        Paper & Country \\ \hline
        Single country studies &  \\ \hline
        \cite{cornwell1996carbon} & Australia \\ 
\cite{eisner2021distributional} & Austria \\ 
\cite{vandyck2014distributional} & Belgium \\ 
\cite{moz2021winners} & Brazil \\ 
\cite{hamilton1994simulating} & Canada \\ 
\cite{winter2023carbon} & Canada \\ 
\cite{dissou2014can} & Canada \\ 
\cite{mardones2020economic} & Chile \\ 
\cite{mardones2024contribution} & Chile \\ 
\cite{zhang2019effects} & China \\ 
\cite{jiang2017analyzing} & China \\ 
\cite{li2020welfare} & China \\ 
\cite{brenner2007chinese} & China \\ 
\cite{jiang2014distributional} & China (Shanghai) \\ 
\cite{wier2005co2} & Denmark \\ 
\cite{wier2003environmental} & Denmark \\ 
\cite{christis2019detailed} & Flanders \\ 
\cite{ravigne2022fair} & France \\ 
\cite{berry2019distributional} & France \\ 
\cite{douenne2020vertical} & France \\ 
\cite{bourgeois2021lump} & France \\ 
\cite{giraudet2021policies} & France \\ 
\cite{semet2024coordinating} & France \\ 
\cite{jacobs2022distributional} & Germany \\ 
\cite{tobben2023unequal} & Germany \\ 
\cite{bach2002effects} & Germany \\ 
\cite{datta2010incidence} & India \\ 
\cite{yusuf2015distributional} & Indonesia \\ 
\cite{o1997carbon} & Ireland \\ 
\cite{tovar2023benefits} & Ireland \\ 
\cite{hynes2009spatial} & Ireland \\ 
\cite{reanos2021fuel} & Ireland \\ 
\cite{reanos2022measuring} & Ireland \\ 
\cite{callan2009distributional} & Ireland \\ 
\cite{verde2009distributional} & Ireland \\ 
\cite{reanos2022distributional} & Ireland \\ 
\cite{campagnolo2022distributional} & Italy \\ 
\cite{tiezzi2005welfare} & Italy \\ 
\cite{renner2018household} & Mexico \\ 
\cite{rosas2017distributional} & Mexico \\ 
\cite{renner2018household} & Mexico \\ 
\cite{kerkhof2008taxation} & Netherlands  \\ 
\cite{creedy2006carbon} & New zealand \\ 
\cite{malerba2021mitigating} & Peru \\ 
\cite{Antosiewicz22} & Poland \\ 
\cite{labandeira1999combining} & Spain \\ 
\cite{brannlund2004carbon} & Sweden \\ 
\cite{landis2019cost} & Switzerland \\ 
\cite{shei2024distributional} & Taiwan  \\ 
\cite{saelim2019carbon} & Thailand \\ 
\cite{saelim2019carbon2} & Thailand \\ 
\cite{phungrassami2019fossil} & Thailand \\ 
\cite{symons1994carbon} & UK \\ 
\cite{bardsley2017something} & UK \\ 
\cite{grainger2010pays} & USA \\ 
\cite{sager2019income} & USA \\ 
\cite{rausch2016household} & USA \\ 
\cite{williams2014initial} & USA \\ 
\cite{rausch2011distributional} & USA \\ 
\cite{hassett2009incidence} & USA \\
\cite{mathur2014distributional} & USA \\
\cite{fremstad2019impact} & USA \\ \hline
Multi-country studies & \\ \hline
\cite{jaccard2021energy} & 28 EU (pooled) \\ 
\cite{buchs21} & 23 EU (pooled) \\ 
\cite{feindt2021understanding} & 23 EU \\ 
\cite{barker1998equity} & 11 EU \\ 
\cite{vandyck2021climate} & 11 EU \\
\cite{rub2024inequality} & 7 EU \\ 
\cite{linden2024manyfaces} & 6 EU\\ 
\cite{symons2002distributional} & 5 EU \\ 
\cite{dorband2019poverty} & 87 low- and middle-income countries  \\
\cite{missbach2024cash} & 16 Latin American and Caribbean \\ 
\cite{vogt2019cash} & 16 Latin American and Caribbean \\ 
\cite{steckel2021distributional} & 8 Asian \\
\cite{chepeliev2021distributional} & Global (23 regions) \\ \hline\hline
\end{longtable}

\newpage

\begin{figure}[ht]
	\centering
		\caption{Study selection process.}

		\copyrightbox[]{\includegraphics[width=1\textwidth]{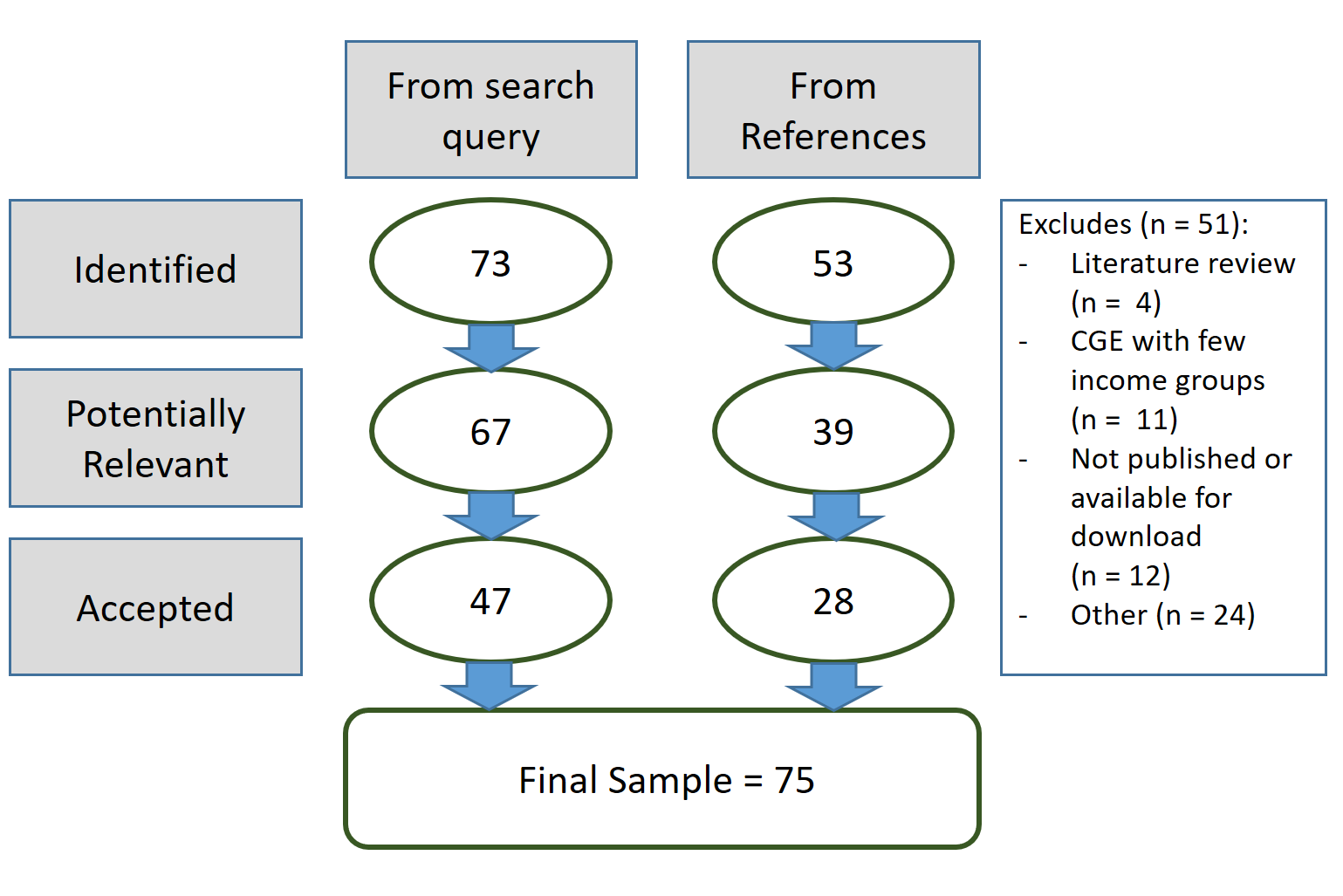}}%
		{}
		
	\label{FIG:Framework}
\end{figure}

\begin{figure}[ht]
	\centering

				\caption{Single-country sample overview.}

		\copyrightbox[]{\includegraphics[width=1\textwidth]{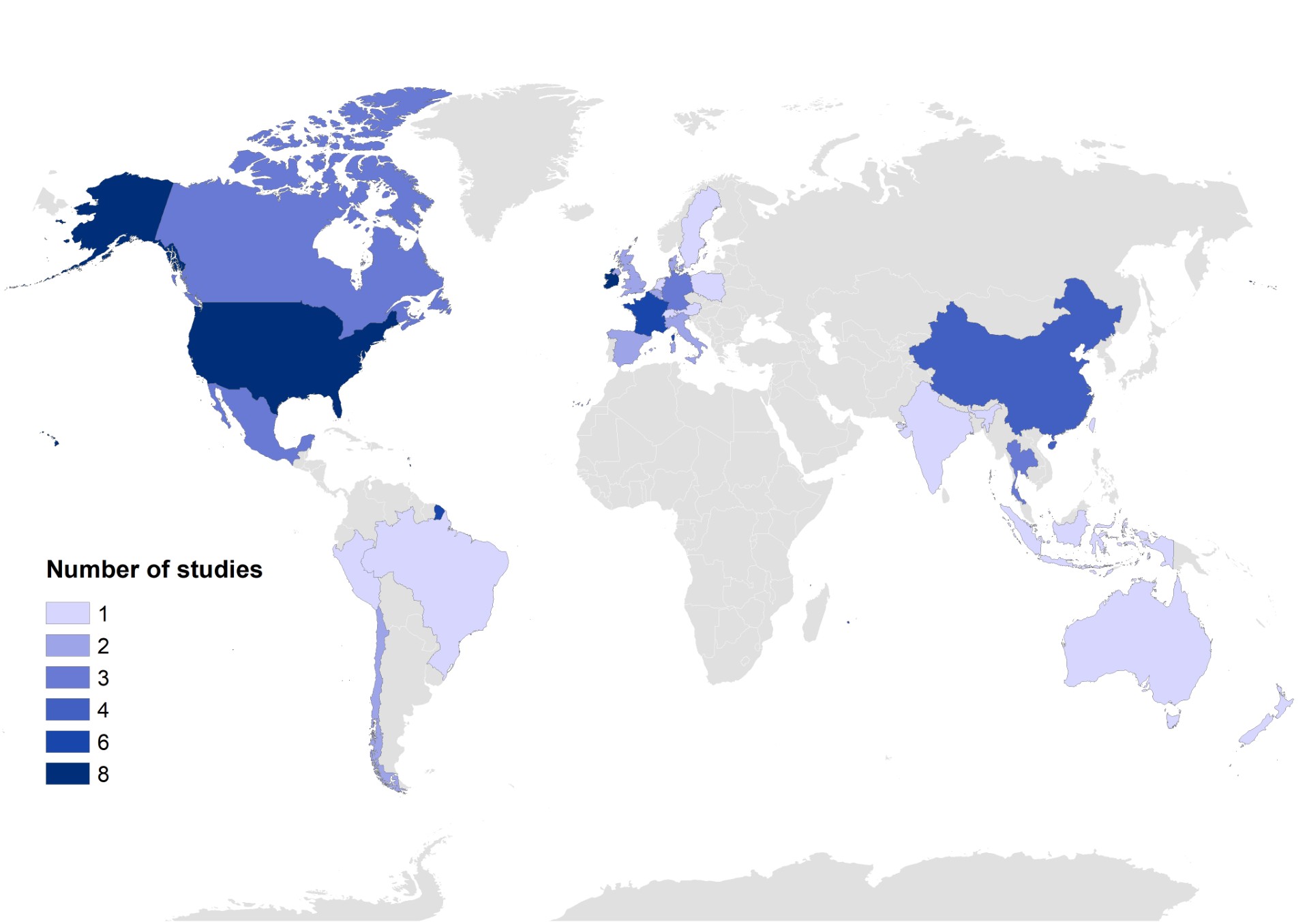}}%
		{}
 	\caption*{\footnotesize Country list: AT AU BE BR CA CH CL CN DE DK ES FR ID IE IN IT MX NL NZ PE PL SE TH TW UK US.} 


	\label{FIG:mapSingle}
\end{figure}

\begin{figure}[ht]
	\centering
				\caption{Number of studies by tax coverage.}

		\copyrightbox[]{\includegraphics[width=0.8\textwidth]{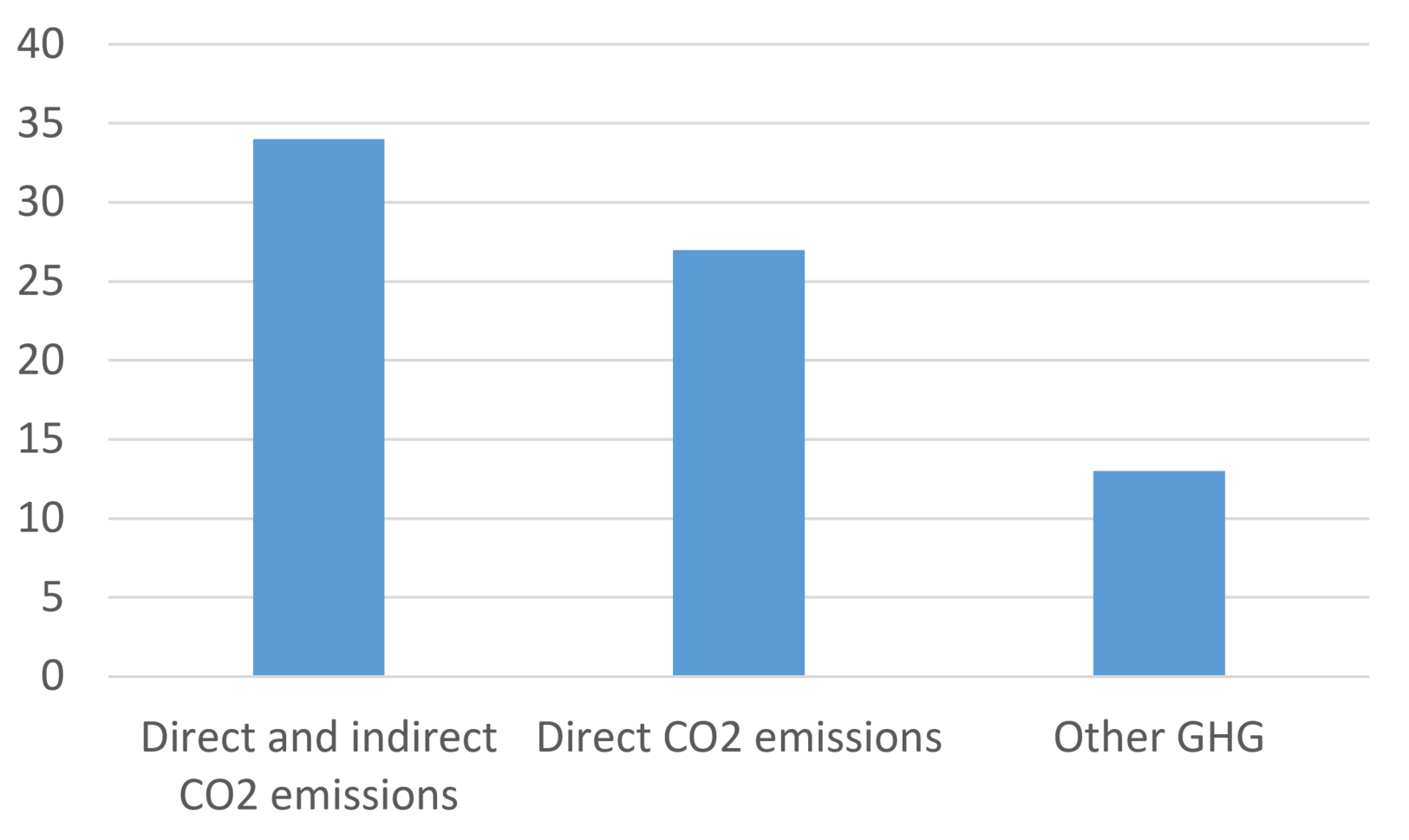}}%
		{}
	\label{FIG:coverage}
\end{figure}

\begin{figure}[ht]
	\centering
				\caption{Counts of simulated revenue recycling schemes across studies.}

		\subfloat[]{
		\copyrightbox[]{\includegraphics[width=0.8\textwidth]{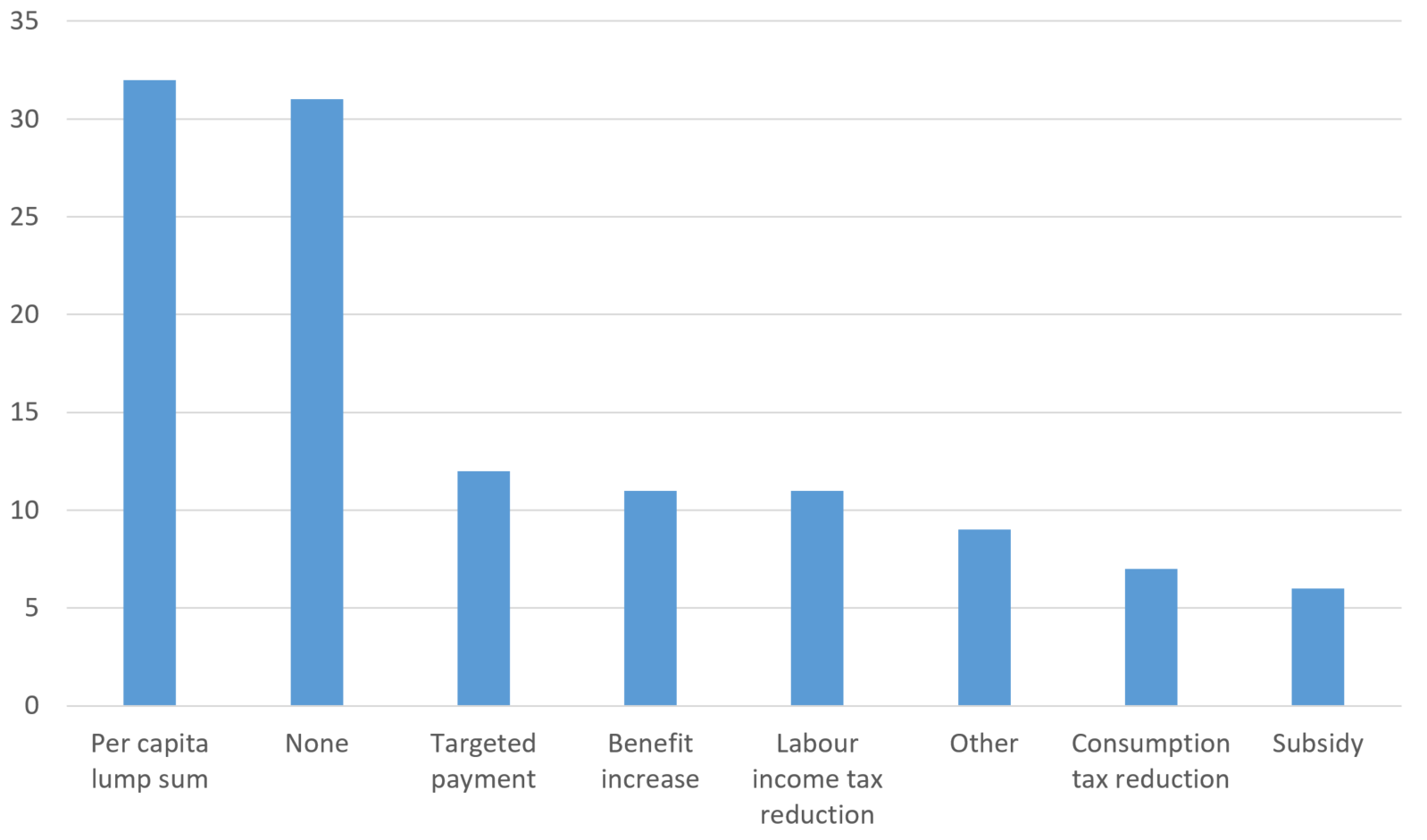}}%
		{}
		}%
	\label{FIG:revrecyOptions}
\end{figure}

\begin{figure}[ht]
	\centering
		\caption{Use of welfare concepts.}

		\subfloat[Use of welfare concepts across studies. \label{fig:fY}]{
		\copyrightbox[b]{\includegraphics[width=0.44\textwidth]{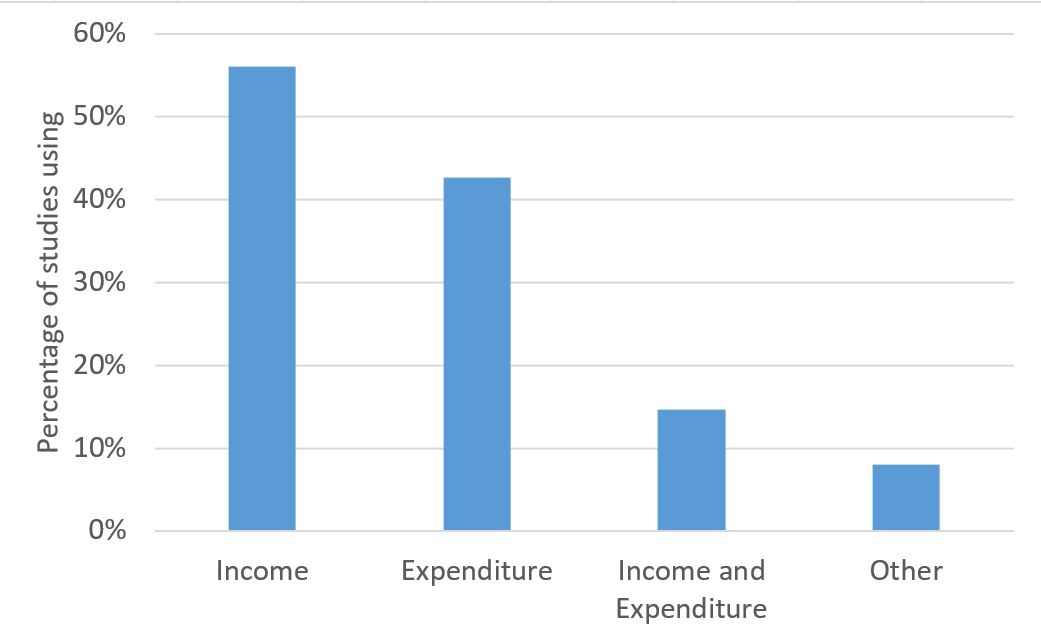}}%
		{}
		}%
		\qquad
		\subfloat[Use of welfare concepts across time. \label{fig:fX}]{
		\copyrightbox[b]{\includegraphics[width=0.44\textwidth]{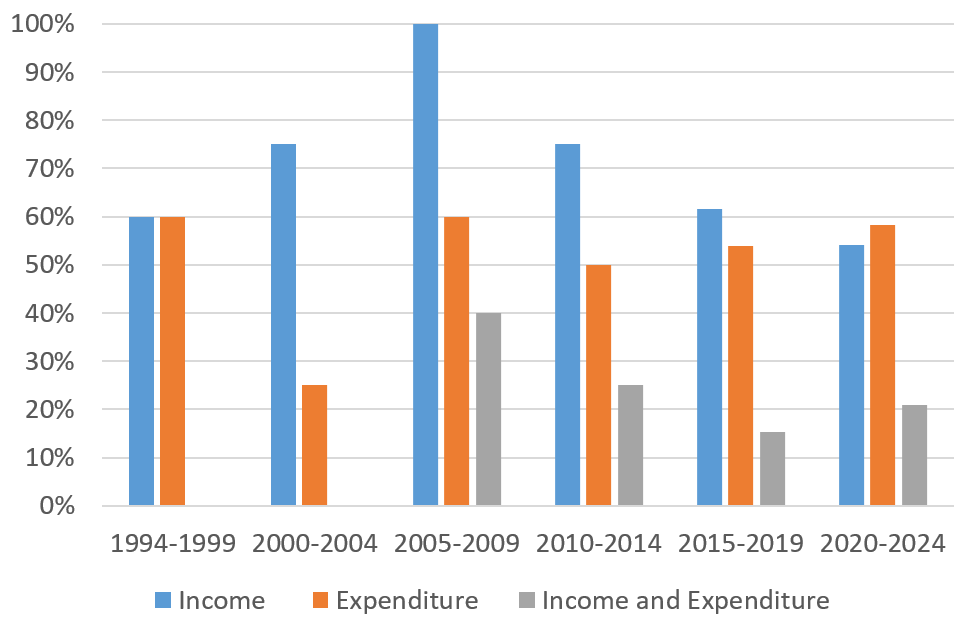}}%
		{}
		}%
	\label{FIG:incConcepts}
\end{figure}

\begin{figure}[ht]
	\centering
				\caption{Counts of horizontal inequalities investigated by type across studies.}

	\copyrightbox[]{\includegraphics[width=0.9\textwidth]{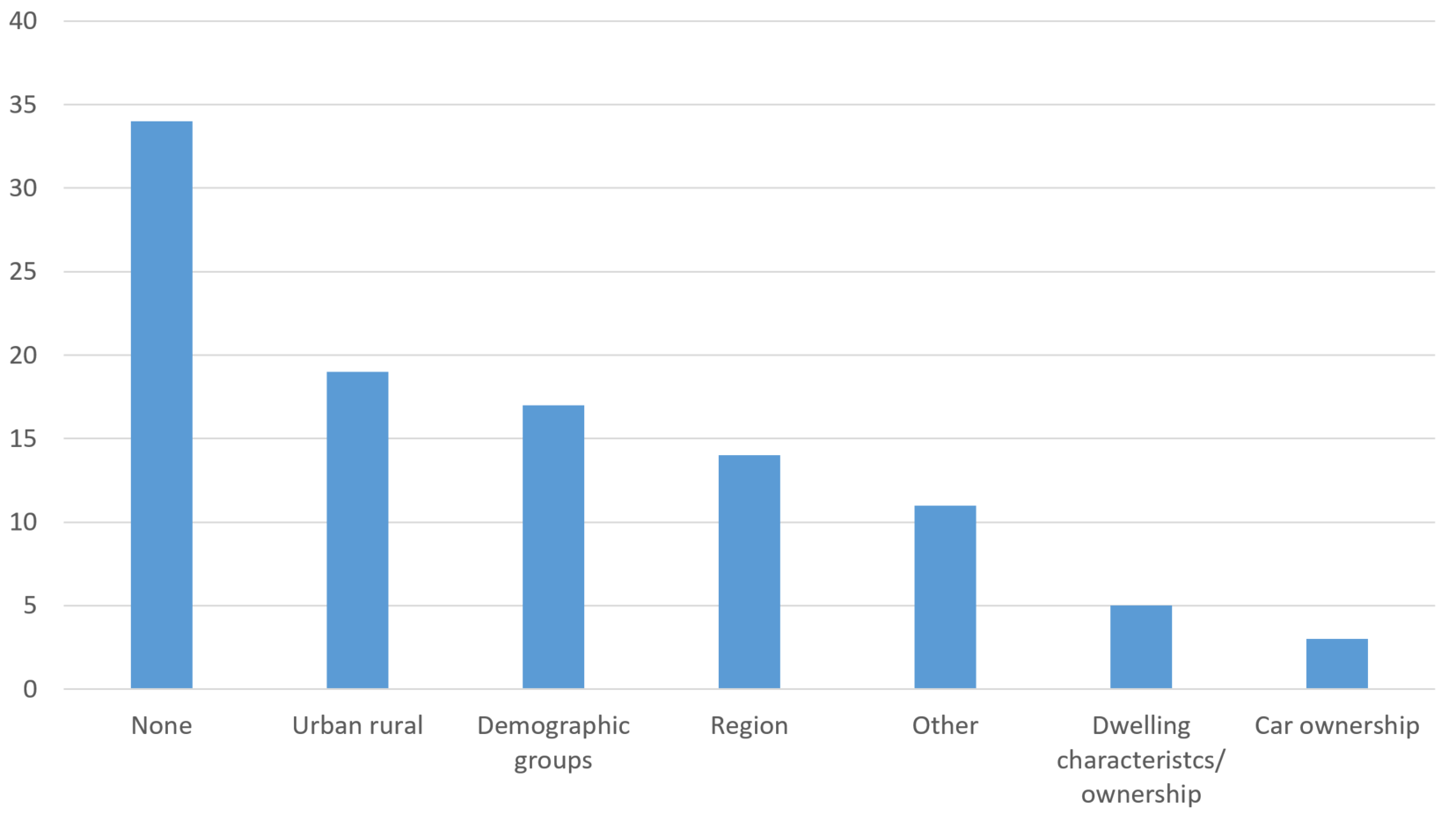}}%
		{}
	\label{FIG:horizontal}
\end{figure}

\newpage

\begin{longtable}{ l c c c c c }
\caption{Descriptive statistics - All estimates} \label{descriptive} \\

\hline
Variable & Count &  Mean & Std. Dev. & Min. & Max. \\    \hline
\endfirsthead
Regressive    & 216  &   .620  &    .486   &  0   &  1  \\
Progressive   &   216 &  .259  &  .439  &   0  &  1  \\
Neutral   &  216 & .111  &  .315   &    0    &   1  \\
HBS Data year   & 214  & 2010  &  7.68  &    1986   & 2021  \\
IO Data year  &   166 & 2010  & 5.59  &   1984  & 2016  \\
IO &  217   &  .857 &    .350  &      0      &      1  \\
MRIO   &  217  & .622 & .486   &  0  &  1  \\
GTAP  &   217  &  .475 & .501  &    0   & 1  \\
GE     &   217  &  .281 &  .451 &   0    &     1  \\
Demand-sided  &  217 & .604 &  .490 &  0  &  1  \\
Life-time income  &  217  &  .424 &     .500    &  0 &  1  \\
Exemptions &    217 & .092 &   .290   & 0   &   1  \\
Imported emissions & 217 & .341 &    .475 &   0   & 1  \\
Other GHG  &  217 & .124 &  .331  &  0 & 1  \\
Other Measure &    217 & .295 &  .457 & 0   &  1  \\
Behav. Adjusted Welfare   &  217  & .240 &    .428 &   0    &  1  \\
Pov. Gap (\$6.85/day)      & 206 & 6.66 &    11.71 &    0  &  61.1  \\
Gini  &  200  & 36.2 &    7.43 & 23   & 63.4  \\
tCO2 per capita  &   210  &  6.72  &  4.50 &    .340 &   29.3   \\
Renewable energy share &    210  &   28.1 &    22.7 &       0  &   99.9  \\
Low income  & 217   &  .230  &  .422 &        0    & 1  \\
Middle income   &    217  &   .373 &   .485 &    0  & 1  \\
High income    &  217 & .396 &  .490 &      0  & 1  \\
Rural pop. share &   210   & 29.3 &  14.9  &   2.12 &  71.1   \\
Low density      & 217  & .267 &   .443 &     0  &   1  \\
Medium density  & 217  & .475 &  .501  &   0    &  1  \\
High density    &   217  &  .258 &  .439    &  0     &   1  \\

\underline{Region \& income level}      & & & \\ 
Eastern Europe \& low inc. &  215  &  .005 &    .068 &         0     &      1 \\

Eastern Europe \& middle inc. &  215  &  .135 &    .342     &     0     &      1 \\

Eastern Europe \& high inc. &  215  &  .023 &    .151     &     0      &     1 \\

Western Europe \& low inc. & 215   & .019 &    .135      &   0       &    1 \\

Western Europe \& middle inc. & 215  &  .107  &   .310  &       0     &      1 \\

Western Europe \& high inc. &   215 &   .302 &   .460    &      0     &      1 \\

South East Asia \& low inc. &   215  &  .051 &  .221     &      0    &       1 \\

South East Asia \& middle inc. & 215  &  .019 &    .135 &        0      &    1 \\

South East Asia \& high inc. &  215  &  .005  &   .068      &    0    &       1 \\

Other Asia \& low inc.& 215 &   .033 &  .178    &     0       &    1 \\

Other Asia \& middle inc.&  215   & .019 &    .135     &      0    &       1 \\

Other \& low inc.& 215  &  .009 &   .096    &      0      &     1 \\

Other \& middle inc. &  215  &  .014 &  .118     &      0       &    1 \\

Other \& high inc. & 215 &   .005 &  .0682   &     0     &      1 \\

South America \& low inc. &  215  &  .116 &   .321     &      0      &     1  \\

South America \& middle inc. & 215  &   .056 &    .230   &      0      &     1 \\

North America \& middle inc. & 215  &   .028 & .165    &      0     &      1  \\

North America \& high inc.  & 215  &   .056  &  .230  &       0       &    1 \\ 

\underline{Region \& income level (fixed effects)}   & & & \\ 
Europe \& low inc.   &    215  &  .023  &  .151   &   0  &      1 \\
Europe \& middle inc.   &    215  &  .2418 &   .429 &         0    &   1 \\
Europe \& high inc.   &    215   & .326  & .470   &   0    &   1 \\
Asia, Africa, Austr. \& low inc.  &    215   & .093 &   .291 &   0     &     1 \\
Asia, Africa, Austr. \& middle inc.   &    215   & .051 &  .221  &   0   &     1 \\
Asia, Africa, Austr.\& high inc.   &    215   & .009 &   .096 &    0     &    1 \\
Americas \& low inc.   &   215   & .116 &  .321   &  0   &  1 \\
Americas \& middle inc.    &  215   & .084  &  .278   &     0  &   1 \\
Americas \& high inc.   &   215   &  .056  &  .230   &    0  &   1  \\\hline
\end{longtable}

\clearpage

\begin{figure}[ht]
	\centering
				\caption{Number of studies per year.}

		\copyrightbox[]{\includegraphics[width=0.8\textwidth]{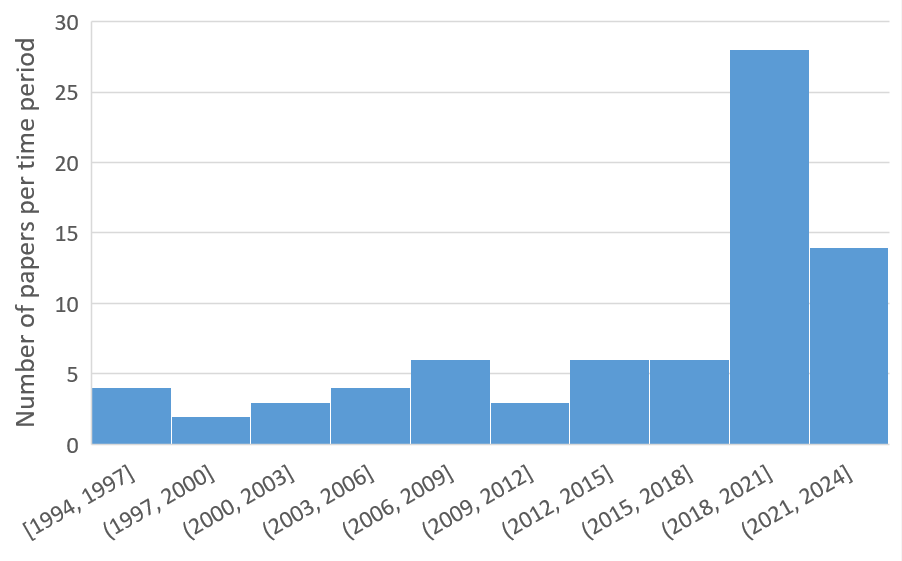}}%
		{}

	\label{FIG:peryear}
\end{figure}

\begin{figure}[ht]
	\centering
				\caption{Cross-country comparative sample overview.}

		\copyrightbox[]{\includegraphics[width=0.7\textwidth]{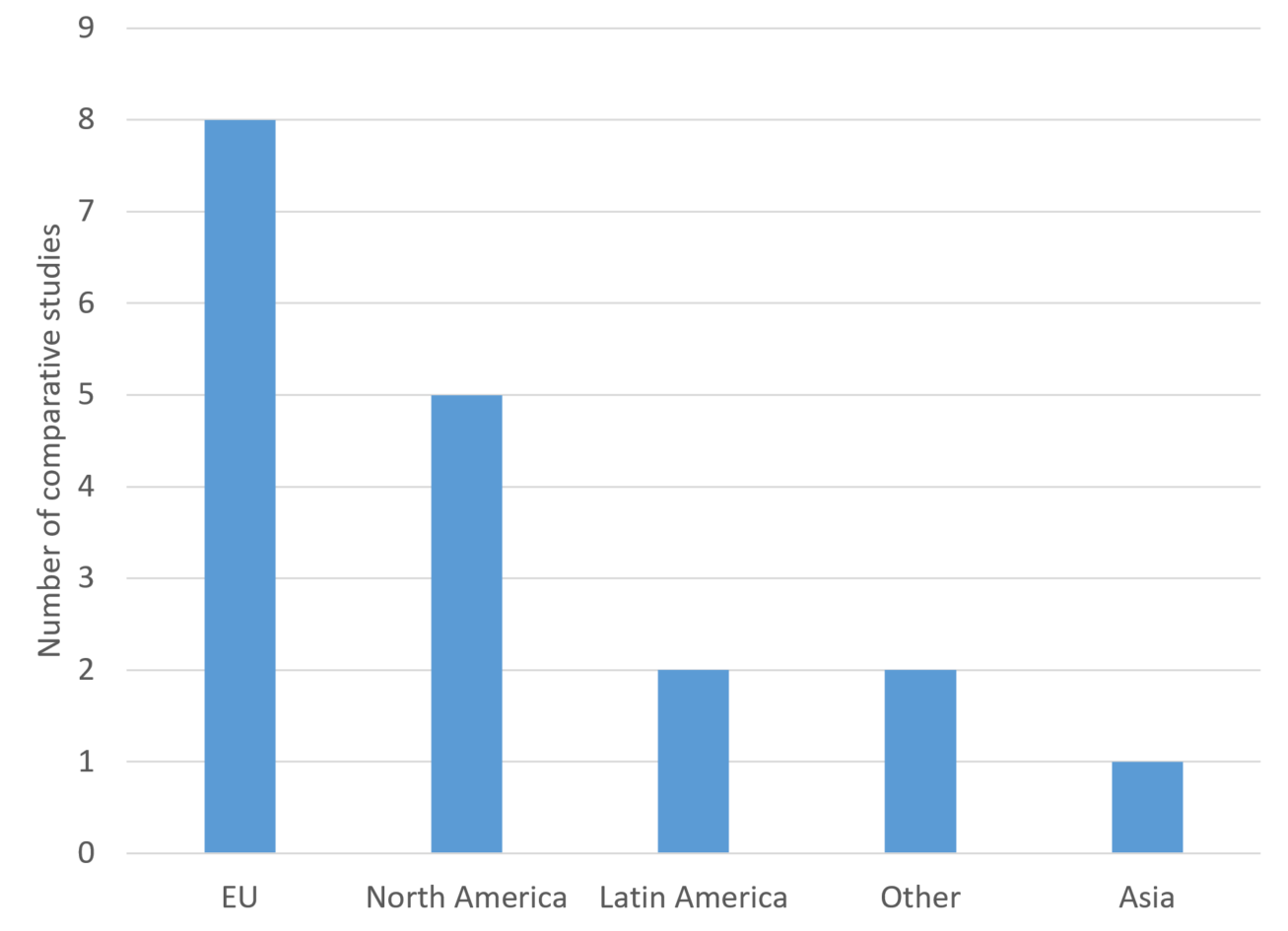}}%
		{}
		\caption*{\footnotesize North America includes comparative countries across states, regions, and provinces. Countries and regions in the global study by \cite{chepeliev2021distributional} are not included in this figure.}

	\label{FIG:multicountry}
\end{figure}

\bibliographystyle{apalike}
\bibliography{sample}

\end{document}